\documentclass[acmsmall]{acmart}
\AtBeginDocument{%
  }

\setcopyright{acmlicensed}
\copyrightyear{2026}
\acmYear{2026}
\acmDOI{Preprint Submitted to CSUR}

\acmJournal{CSUR}
\acmVolume{XX}
\acmNumber{X}
\acmArticle{111}
\acmMonth{8}

\usepackage{pifont}  
\newcommand{\cmark}{\checkmark}
\newcommand{\xmark}{\ding{55}}
\usepackage{balance}
\usepackage{enumitem}
\usepackage[export]{adjustbox}  
\usepackage{wrapfig}

\usepackage{comment}
\usepackage{amsmath}
\usepackage{graphicx}
\usepackage{subcaption}
\usepackage{float}
\usepackage{caption}
\usepackage{multirow}
\usepackage{booktabs}
\usepackage{adjustbox}
\usepackage{placeins}
\usepackage[table,xcdraw]{xcolor} 
\usepackage{algorithm}
\usepackage{algpseudocode}
\newcommand{\bolditalic}[1]{\textbf{\textit{#1}}}
\usepackage{ragged2e}
\usepackage{adjustbox}
\usepackage[flushleft]{threeparttable}  
\usepackage{array}
\usepackage{tabularx}
\newcolumntype{P}[1]{>{\centering\arraybackslash}p{#1}}
\newcolumntype{M}[1]{>{\centering\arraybackslash}m{#1}}
\newcolumntype{L}[1]{>{\centering\arraybackslash}l{#1}}
\newcolumntype{Z}{>{\centering\let\newline\\\arraybackslash\hspace{0pt}}X}

\newcolumntype{Y}{>{\raggedright\let\newline\\\arraybackslash\hspace{0pt}}X}
\usepackage{colortbl}
\usepackage{color,soul}
\usepackage{hhline}
\usepackage{makecell}
\usepackage{xurl}
\usepackage{hyperref}
\hypersetup{
    breaklinks=true,
    colorlinks=true,
    linkcolor=blue,
    citecolor=blue,
    urlcolor=blue
}

\usepackage[figuresright]{rotating}

\begin{document}

\title{Impact of Intelligent Technologies on IoV Security: Integrating Edge Computing and AI}

\author{Awais Bilal}
\authornote{Both authors contributed equally.}
\email{awaisbilal@bit.edu.cn}
\author{Kashif Sharif}
\authornotemark[1]
\authornote{Corresponsing Author. \\This work was supported by Beijing Natural Science Foundation (IS23056).}
\orcid{1234-5678-9012}
\email{kashif@bit.edu.cn}
\affiliation{%
  \department{School of Computer Science and Technology}
  \institution{Beijing Institute of Technology}
  \city{Beijing}
  \country{China}
}

\author{Liehuang Zhu}
\email{liehuangz@bit.edu.cn}
\author{Chang Xu}
\email{xchang@bit.edu.cn}
\affiliation{%
  \department{School of Cyberspace Science and Technology}
  \institution{Beijing Institute of Technology}
  \city{Beijing}
  \country{China}}

\author{Fan Li}
\email{fli@bit.edu.cn}
\author{Sadaf Bukhari}
\email{sadaf@bit.edu.cn}
\affiliation{%
  \department{School of Computer Science and Technology}
  \institution{Beijing Institute of Technology}
  \city{Beijing}
  \country{China}
}

\author{Sujit Biswas}
\affiliation{%
  \department{FinTech Department of Computer Science, School of Science and Technology}
  \institution{CITY, University of London}
  \city{London}
  \country{UK}}
\email{sujit.biswas@city.ac.uk}

\renewcommand{\shortauthors}{Bilal et al.}



\begin{abstract}
  The rapid development and integration of intelligent technologies in the Internet of Vehicles (IoV) have revolutionized transportation systems by enhancing connectivity, automation, and safety. However, the complexity and connectivity of IoV networks also introduce security challenges, including data privacy concerns, cyber threats, and system vulnerabilities. This paper surveys the role of Edge Computing (EC), Machine Learning (ML), and Deep Learning (DL) in strengthening IoV security frameworks. It examines the synergy between these technologies, highlighting their individual capabilities and their collective impact on enhancing threat detection, response times, and adaptive security. Through real world case studies and practical deployments, we demonstrate how EC, ML, and DL are currently improving security and operational efficiency in IoV systems. The paper also identifies key research gaps and future directions for further advancements in IoV security, including the need for scalable, privacy preserving solutions and robust defense mechanisms against emerging cyber threats. By integrating EC, ML, and DL, this work lays the groundwork for developing adaptive, efficient, and resilient IoV security infrastructures capable of addressing evolving challenges in the transportation ecosystem.
\end{abstract}




\maketitle

\section{Introduction}\label{sec:introduction}
The Internet of Vehicles (IoV) integrates vehicles, roadside units, and cloud platforms into a unified, cyber physical transportation network, enabling real time data exchange for traffic management, V2I communication, and automated driving services~\cite{Wang2021Green}.  However, pervasive connectivity also broadens the attack surface: compromised sensors, spoofed messages, or hijacked control channels can breach privacy, disrupt services, and endanger lives~\cite{Mollah2020Blockchain}.

\begin{wrapfigure}{r}{0.5\textwidth} 
    \centering
    \includegraphics[width=0.48\textwidth]{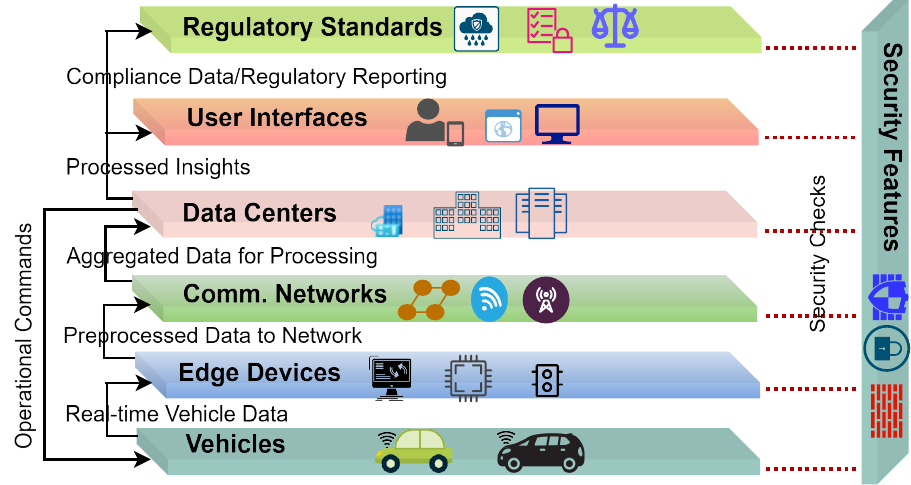}
    \caption{Data flow and interactions within the IoV framework: Integrating components with security measures.}
    \label{fig:IoVEcoSystem}
    \vspace{-10pt} 
\end{wrapfigure}

Conventional security solutions, firewalls, signature based Intrusion Detections Systems (IDS), and centralised analysis, struggle with IoV’s stringent latency requirements (sub-10 ms), high mobility, and device heterogeneity.  Edge Computing (EC) mitigates these constraints by processing data locally, cutting round trip delays and reducing backbone exposure~\cite{Grover2021Edge}.  Machine Learning (ML) adapts to evolving attack patterns via predictive classifiers and clustering~\cite{Ning2020Joint,Chen2021An}, while Deep Learning (DL) uncovers subtle, high dimensional anomalies that shallow models miss~\cite{Chen2019Deep,Dai2019Artificial}.

Each technology alone reveals gaps: EC lacks deep analytics, ML falters on zero day exploits, and DL can overwhelm resource limited nodes.  Their \emph{synergistic integration}, e.g., edge hosted ML filters that pre-screen data for downstream DL inference, plus federated updates that protect raw data, promises a real time, scalable, and privacy preserving IoV security fabric~\cite{Xu2023Safe:}.

The Figure~\ref{fig:IoVEcoSystem}, offers a comprehensive view of the dynamic interactions and data flow among all technologies within the IoV framework. This figure illustrates the essential roles of various IoV components, showing how data traverses the system from vehicles to edge devices and then to central data hubs, finally reaching end user interfaces, all underpinned by robust security protocols and standards. The synergy between these technologies is vital for strengthening IoV security.

\bolditalic{Scope of this Survey:}~This survey examines the latest advancements in EC, ML, and DL with a specific focus on their role in IoV security, covering research published between 2019 and 2024. Foundational works are referenced where necessary to provide historical context and highlight the evolution of artificial intelligence driven cybersecurity in IoV. This study not only analyzes the theoretical and practical integration of edge intelligence and AI-based security mechanisms but also presents real world applications through case studies, illustrating how these technologies enhance threat detection, intrusion prevention, and data privacy in IoV. Furthermore, this survey examines emerging technological trends that are expected to reshape IoV security frameworks, including quantum computing, AI-driven autonomous security models, and next generation networks. By identifying current challenges and future directions, this work provides a structured foundation for researchers and practitioners seeking to advance scalable, adaptive, and privacy preserving security solutions for IoV ecosystems.

\bolditalic{Key Research Questions:}~This survey explores the integration of EC, ML, and DL in securing IoV systems by addressing the following research questions:

\begin{enumerate}
    \item What are the distinct vulnerabilities and threat vectors present within different IoV architectures and connectivity models?
    \item How does the incorporation of EC influence latency reduction, scalability, and real time security enforcement in IoV networks?
    \item In what ways do ML and DL techniques advance the detection, classification, and mitigation of security threats compared to traditional approaches?
    \item How do ML and DL models enhance pattern recognition and anomaly detection capabilities within complex, high dimensional IoV data streams?
    \item What synergistic benefits emerge from integrating EC with ML and DL technologies in designing adaptive and resilient IoV security frameworks?
    \item How does the adoption of emerging technologies affect regulatory compliance and the evolution of industry standards within the IoV security domain?
    \item What operational and technological challenges persist in deploying integrated IoV security frameworks, particularly concerning interoperability, system maintenance, and scalability?
    \item What future directions and emerging research areas are poised to drive advancements in AI-driven IoV security frameworks?
\end{enumerate}

\begin{figure*}[!t]
    \centering
    \includegraphics[width=0.95\textwidth]{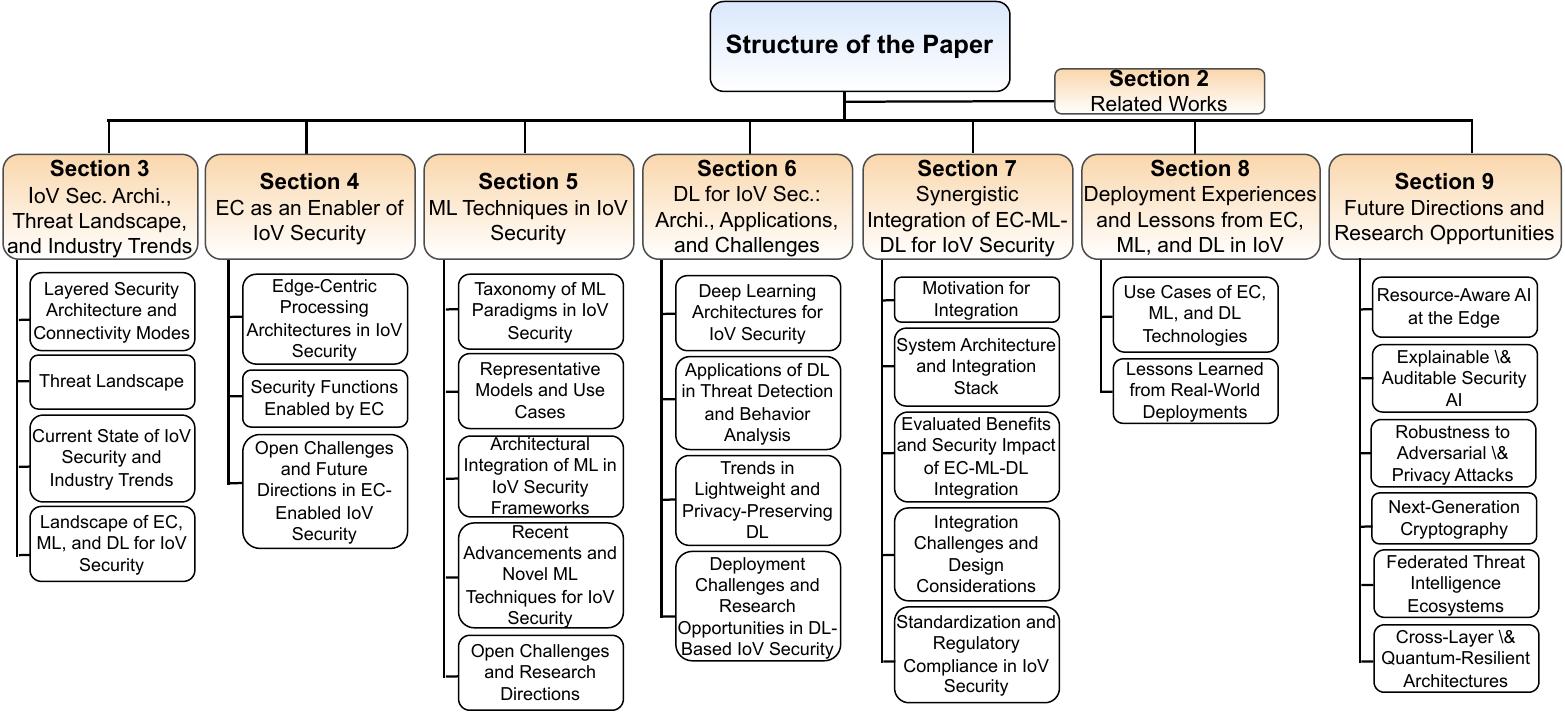}  
    \caption{Overview of the paper structure.}
    \label{fig:structure-review-paper}
\end{figure*}

\bolditalic{Structure of the Paper:}~This survey explores the integration of advanced technologies in IoV security. Figure~\ref{fig:structure-review-paper}, outlines the organization. Section~\ref{relatedworks}, \emph{Related Works}, reviews prior surveys and positions this study’s contributions. Section~\ref{sec:iov_sec_arch_industory_trends}, \emph{IoV Security Architecture, Threat Landscape, and Industry Trends}, discusses layered models, threat evolution, and industry responses. Section~\ref{sec:role-edge-computing}, \emph{Role of EC in IoV Security}, highlights EC’s role in real time defense. Sections~\ref{sec:ml-techniques} and~\ref{sec:dl-iov-security}, \emph{ML} and \emph{DL Techniques}, examine AI approaches for adaptive threat mitigation. Section~\ref{synergistic-effects}, \emph{Synergistic Effects of Combining Technologies}, assesses integrated EC-ML-DL frameworks. Section~\ref{case-studies}, \emph{Case Studies}, presents real world deployments, and Section~\ref{PredictionsandResearchOpportunities}, \emph{Future Directions and Research Opportunities}, identifies open challenges. Section~\ref{sec:Conclusion} concludes with a synthesis of insights.

\begin{table*}[!t]
\centering
\begin{threeparttable}
\caption{Comparison of existing surveys on IoV security and research gaps analyzed in this work.}
\label{tab:existingSurveysAndOurContribution}
\fontsize{6pt}{7.2pt}\selectfont
\begin{tabularx}{0.98\linewidth}{
  >{\hsize=1.0\hsize}Y 
  >{\hsize=0.3\hsize\centering\arraybackslash}Y 
  >{\hsize=0.3\hsize\centering\arraybackslash}Y 
  >{\hsize=0.3\hsize\centering\arraybackslash}Y 
  >{\hsize=1.6\hsize}Y 
  >{\hsize=1.6\hsize}Y 
  >{\hsize=1.9\hsize}Y 
}

\toprule
\textbf{Ref (Year)} & \textbf{EC} & \textbf{ML} & \textbf{DL} & \textbf{Focus} & \textbf{Gaps} & \textbf{Improvements in This Work} \\
\hline
\cite{tang2019future}$\oplus$ (2019) & \xmark & \cmark & \xmark & ML in 5G/6G IoV & Missing DL + EC & Unified ML + DL + EC \\ \hline
\cite{olowononi2020resilient}$\ast$ (2020) & \xmark & \cmark & \xmark & Resilience in IoV & Privacy; EC & Privacy-preserving ML + EC \\ \hline
\cite{wang2020convergence}$\ast$ (2020) & \cmark & \xmark & \cmark & EC for DL & Real-time detection; resource mgmt & EC-enhanced ML/DL security \\ \hline
\cite{deng2020edge}$\ddag$ (2020) & \cmark & \xmark & \xmark & EC real-time processing & DL security & EC + DL + ML integration \\ \hline
\cite{qayyum2020securing}$\ast$ (2020) & \xmark & \cmark & \xmark & Adversarial defense & EC for adversarial ML & ML + DL + EC defense \\ \hline
\cite{talpur2021machine}$\ast$ (2021) & \xmark & \cmark & \xmark & ML threat mitigation & EC integration & Edge-augmented ML mitigation \\ \hline
\cite{chiroma2021deep}$\ddag$ (2021) & \xmark & \xmark & \cmark & DL analytics & Security analytics & DL + real-time detection \\ \hline
\cite{murshed2021machine}$\ast$ (2021) & \cmark & \cmark & \xmark & Edge ML processing & Security at edge & Scalable edge ML detection \\ \hline
\cite{djigal2022machine}$\ast$ (2022) & \cmark & \cmark & \cmark & Resource allocation & Security measures & Secure resource allocation \\ \hline
\cite{wang2023blockchain}$\dag$ (2023) & \cmark & \xmark & \xmark & Blockchain integrity & ML/DL detection & ML/DL + Blockchain security \\ \hline
\cite{liu2023crs}$\dag$ (2023) & \cmark & \xmark & \cmark & Privacy-preserving IoV & Scalability & Scalable privacy-preserving framework \\ \hline
\cite{boualouache2023survey}$\ast$ (2023) & \cmark & \cmark & \cmark & 5G-V2X misbehavior & Real-time; scalability & Real-time EC integration \\ \hline
\cite{almehdhar2024deep}$\ast$ (2024) & \xmark & \xmark & \cmark & DL for IDS & Scalability; edge processing & Scalable edge-driven IDS \\ \hline
\textbf{This Work}\newline (2025) & \cmark & \cmark & \cmark & Integrated IoV security & --- & Unified, real-time ML/DL/EC focus \\

\midrule
\end{tabularx}

\begin{tablenotes}[flushleft]\scriptsize
\item ML: Machine Learning; DL: Deep Learning; EC: Edge Computing.
\item $\oplus$: Security-Focused; $\ast$: Survey/Review; $\ddag$: Applied Research; $\dag$: Framework/Model.
\item \cmark = Present, \xmark = Not Present.
\end{tablenotes}
\end{threeparttable}
\end{table*}

\section{Related Works}
\label{relatedworks}
\subsection{Comparison of Existing Works}

Existing surveys and comparative studies on IoV security has explored ML, DL, and EC, but most studies treat these technologies in isolation, lacking a unified or integrated approach. Table~\ref{tab:existingSurveysAndOurContribution} classifies the surveyed literature into four categories: security focused studies, survey/review papers, applied research, and framework/model proposals. This categorization helps distinguish between foundational surveys, practical implementations, and theoretical models, highlighting existing gaps in cross technology integration. Several works have investigated ML and DL based security mechanisms, primarily focusing on threat detection and anomaly classification~\cite{talpur2021machine,qayyum2020securing,almehdhar2024deep}. However, these studies often overlook scalability issues and fail to address how EC can enhance real time processing in resource-constrained vehicular environments. Conversely, research focused on EC highlights its benefits for IoV security, including latency reduction and computational offloading~\cite{deng2020edge,wang2020convergence}. Yet, these approaches generally lack integration with ML/DL models needed for adaptive and intelligent security systems. Privacy preserving techniques such as FL and blockchain have also been examined in recent literature~\cite{liu2023crs,wang2023blockchain}. These works emphasize data integrity and privacy but do not fully address computational efficiency or the challenges of scaling such methods in large vehicular networks.

This survey advances the existing body of work by offering an integrated analysis of EC, ML, and DL technologies in the context of threat detection, scalability, and privacy preserving mechanisms. It further investigates emerging challenges in 5G/6G network environments, AI-driven adaptive security, and quantum-resistant cryptographic approaches, delivering a comprehensive and forward-looking perspective on IoV security frameworks.
Some of the crucial \emph{research gaps} in comparative studies are:

\begin{itemize}
    \item \emph{Fragmented integration.} EC, ML, and DL are often evaluated in isolation; cross layer frameworks that coordinate them remain scarce~\cite{Grover2021Edge}.
    \item \emph{Privacy vs.\ performance.} Certificateless crypto, FL, and blockchain each tackle privacy, but their joint use with edge intelligence is poorly understood~\cite{xie2022efficient,wang2023blockchain}.
    \item \emph{Real time scalability.} Offloading lowers latency~\cite{xu2019edge} yet little work couples it with adaptive ML/DL for large scale, dynamic fleets.
    \item \emph{Next-gen threats.} 5G/6G slicing attacks, adversarial ML, and post quantum cryptography are recognized individually~\cite{tang2019future} but not analysed within an integrated IoV security stack.
\end{itemize}

\subsection{Our Contributions}
\begin{figure*}[!t]
  \centering
  
  \includegraphics[width=0.90\textwidth]{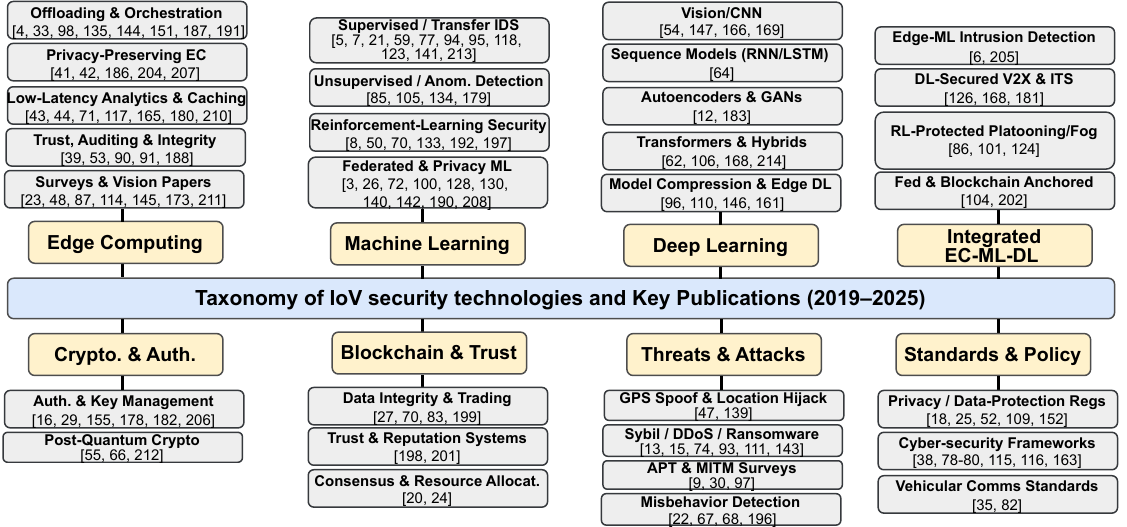}
  \caption{Taxonomy of IoV security technologies and key publications (2019--2025).}
  \label{fig:taxonomy_tree}
\end{figure*}

To address the fragmentation and rapid evolution of IoV security research, this survey makes five key contributions:
\begin{enumerate}
    \item \emph{Comprehensive Taxonomy.} We synthesize over 200 works published between 2019 and 2025 into a unified taxonomy in Figure~\ref{fig:taxonomy_tree}, categorizing EC, ML, DL, and fully integrated EC--ML--DL solutions according to their primary security functions and deployment models.

    \item \emph{In‐Depth Technology Analysis.} For each pillar (EC, ML, DL) we critically examine fundamental methods (e.g., offloading strategies, supervised vs.\ unsupervised learning, CNN/RNN architectures), quantify their performance trade‐offs in latency, accuracy, energy, and privacy, and compare edge native optimizations such as pruning, quantization, and federated updates.
    \item \emph{Real World Case Studies \& Lessons.} We curate a diverse set of large scale deployments, from keyless vehicle authentication and collision avoidance to city scale traffic monitoring, and distill actionable lessons on data governance, scalability, interoperability, continuous adaptation, and user privacy.
    \item \emph{Challenges \& Mitigation Strategies.} Drawing on the literature and our case study insights, we map out critical obstacles, resource constraints, network reliability, regulatory compliance, and propose concrete mitigation tactics, including neuromorphic/event driven inference, blockchain anchored model provenance, and hybrid cryptographic schemes.
    \item \emph{Future Research Roadmap.} We chart a forward looking agenda covering post quantum and homomorphic encryption, explainable and adversarially robust AI, cross layer adaptive security architectures, and 6G enabled V2X trust frameworks, each linked to open questions that will guide the next wave of IoV security innovation.
\end{enumerate}

\begin{figure*}[!t]
    \centering
    \includegraphics[width=\textwidth,%
                     height=0.4\textheight,%
                     keepaspectratio]{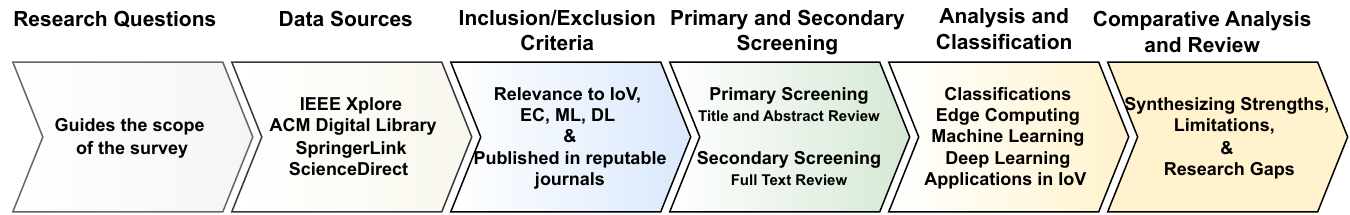}
    \caption{Search methodology for IoV security survey: Systematic data collection, screening, \& analysis.}
    \label{fig:Search_Methodology}
\end{figure*}
\subsection{Literature Review Process}
Figure~\ref{fig:Search_Methodology} depicts our three stage search. We queried IEEE Xplore, ACM DL, SpringerLink, and ScienceDirect with terms such as ``IoV security'', ``edge computing'', ``machine/deep learning in IoV'',  and ``FL''; filters retained peer reviewed work from 2019-2024, plus seminal older papers. Full text screening assessed methodological rigour and real world relevance. Selected studies were coded into themes: EC frameworks, ML based defences, DL anomaly detection, and privacy preserving techniques (FL, blockchain). The process supports a transparent, reproducible evidence base for the survey’s findings.
%
%
%

\begin{wrapfigure}{r}{0.48\textwidth}
  \centering
  \includegraphics[width=\linewidth]{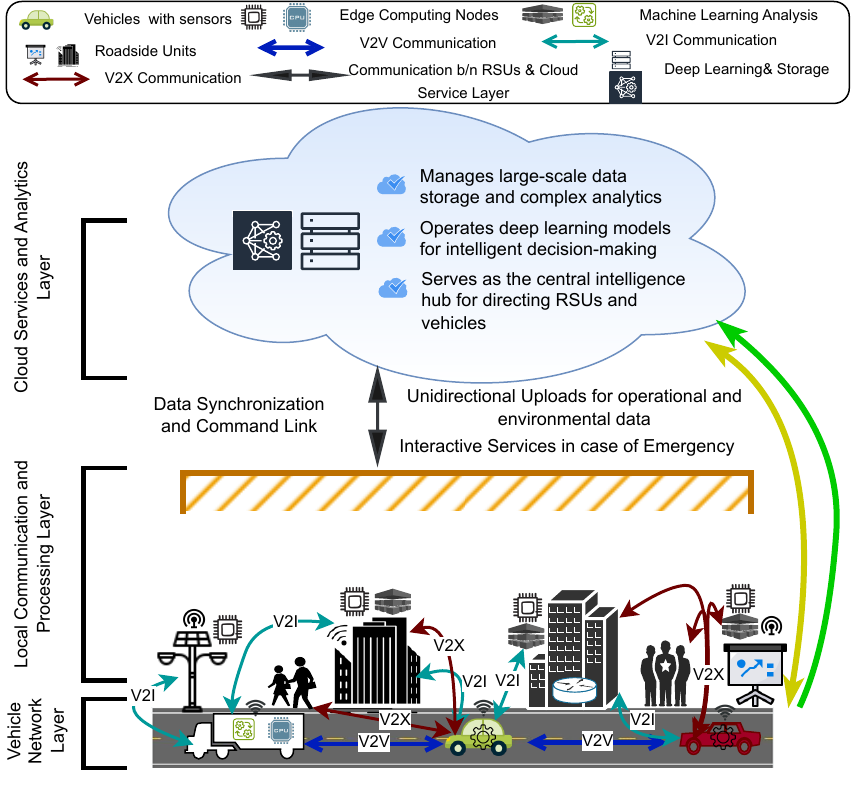}
  \caption{Hierarchical architecture of IoV.}
  \label{fig:IoVArchitecture}
\end{wrapfigure}
\section{IoV Security Architecture, Threat Landscape, and Industry Trends}
\label{sec:iov_sec_arch_industory_trends}

\subsection{Layered Security Architecture and Connectivity Modes}
\label{sec:iov_arch}

The security architecture of the IoV is typically conceptualized as a three-layered model, as illustrated in Figure~\ref{fig:IoVArchitecture}. This layered view facilitates the decomposition of security responsibilities across vehicular, edge, and cloud infrastructures. At the \emph{vehicle network layer}, on-board units (OBUs) are responsible for authenticating V2V communications using short lived public key infrastructure (PKI) credentials. These units also perform frame validation to filter malformed or malicious packets at the point of origin. The \emph{local communication layer}, which operates through RSUs, is tasked with aggregating traffic, applying intrusion detection mechanisms, and enforcing rate limiting to counter volumetric DoS attacks. The uppermost \emph{cloud analytics layer} integrates telemetry from distributed vehicular sources with external threat intelligence feeds. This layer increasingly leverages DL classifiers and FL frameworks to support privacy-preserving threat detection and adaptive response.


\begin{figure}[!b]
  \centering
  \begin{minipage}[t]{0.48\textwidth}
    \centering
    \includegraphics[width=\linewidth]{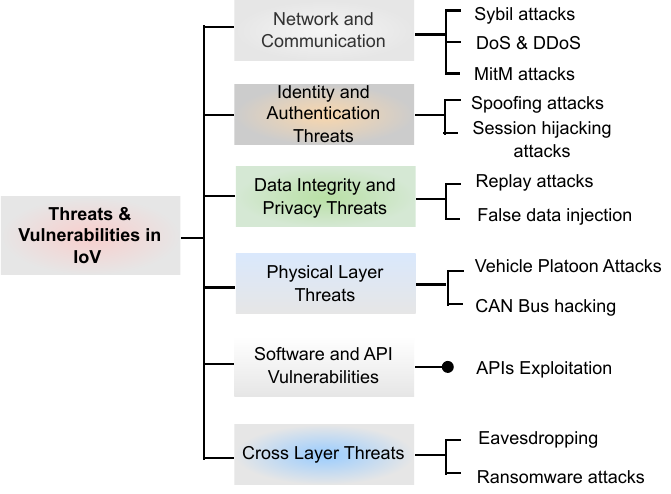}
    \caption{Security threats in the IoV ecosystem.}
    \label{fig:IoVThreatModel}
  \end{minipage}
  \hfill
  \begin{minipage}[t]{0.48\textwidth}
    \centering
    \includegraphics[width=\linewidth]{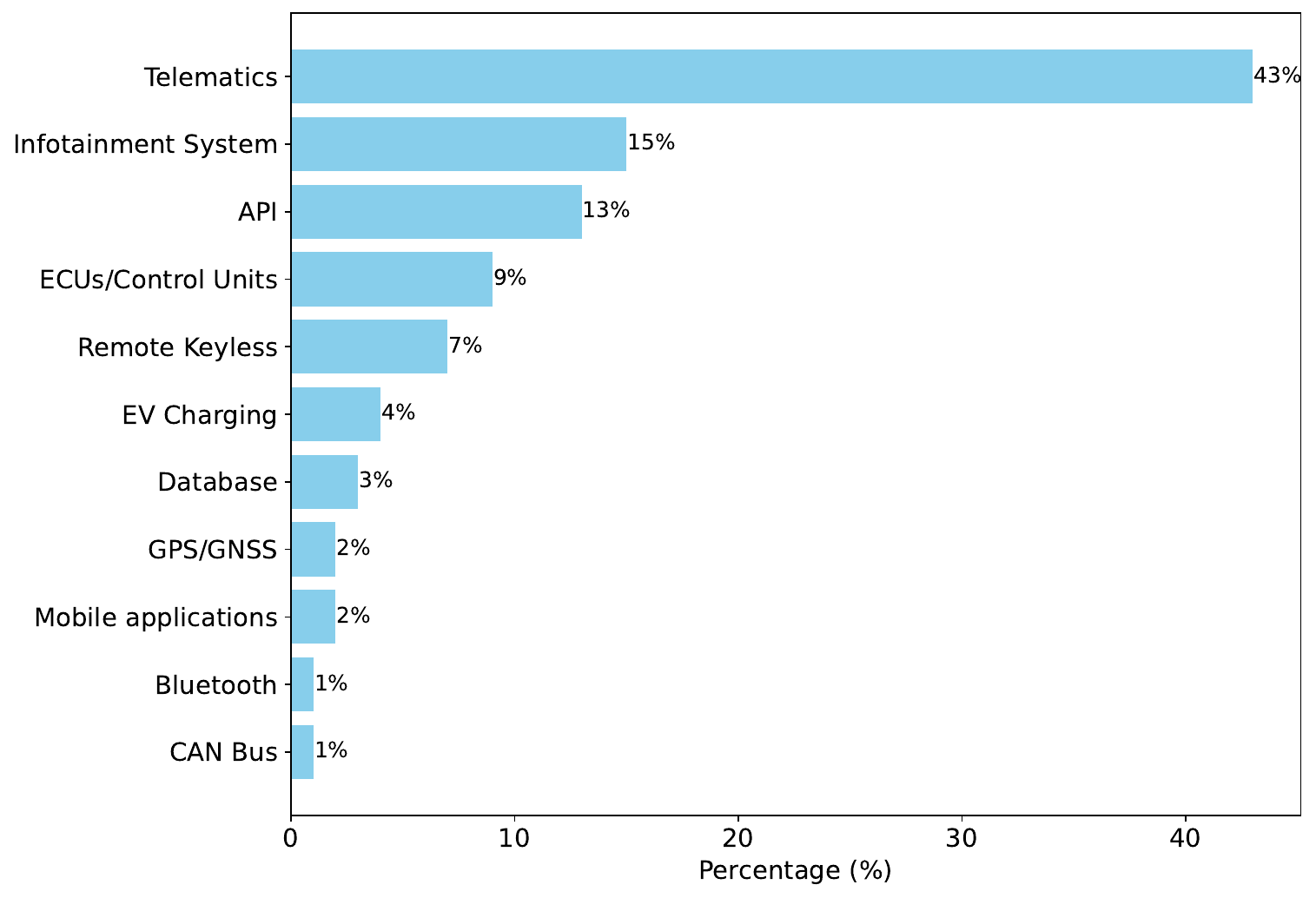}
    \caption{Distribution of different attack vectors in connected vehicles (2023)~\cite{businessresearch2023}.}
    \label{fig:Incidents_by_Attack_vectors}
  \end{minipage}
\end{figure}

\begin{table*}[!t]
\centering
\caption{Summary of unique threats to IoV systems: Types, Methods, and Impacts.}
\fontsize{6pt}{7.2pt}\selectfont
\label{tab:iov_threats}
\begin{tabularx}{\linewidth}{
  >{\hsize=0.7\hsize}Y
  >{\hsize=0.8\hsize}Y
  >{\hsize=1.2\hsize}Y
  >{\hsize=1.3\hsize}Y
}
\toprule
\textbf{Refs} & \textbf{Threat} & \textbf{Examples} & \textbf{Impact} \\
\hline
\cite{Li2021RTED-SD:,Mishra2019Analytical}
& Sybil
& Fake IDs; traffic fabrication
& Network disruption; decision confusion \\
\hline
\cite{Sherazi2019DDoS,Li2021RTED-SD:,Xiao2021Secure}
& DoS/DDoS
& Flooding; botnet floods
& Performance drop; service outage \\
\hline
\cite{Conti2016A,Bagga2021On}
& MitM
& Data injection; response tampering
& Breach of confidentiality; integrity loss \\
\hline
\cite{Chen2019A,Sanders2020Localizing,Dasgupta2021Sensor}
& Spoofing
& GPS/device spoofing
& Misled navigation; safety risk \\
\hline
\cite{Cui2018Privacy-Preserving,Song2020FBIA:}
& Identity theft
& Credential theft; function takeover
& Privacy breach; safety hazard \\
\hline
\cite{garcia2018security,Wazid2019AKM-IoV:,Xu2019An}
& Session hijack
& Route alteration; restricted-access
& Integrity compromise \\
\hline
\cite{Chen2019A}
& API vuln.
& Exploit APIs
& System compromise \\
\hline
\cite{Huang2018Secure,Bajracharya2021Performance}
& Eavesdropping
& Packet sniffing; traffic analysis
& Privacy loss; RSU overload \\
\hline
\cite{Li2021Transfer}
& Timing
& Crypto timing leaks
& Behavior inference; privacy leak \\
\hline
\cite{Chen2019A,Katragadda2020Detecting,Sutrala2020On}
& Replay
& Frame replay; packet delay
& Management disruption \\
\hline
\cite{Aboelwafa2020A,Guo2021Optimal,Yang2021A}
& False data
& Traffic falsification; status spoof
& Wrong decisions; inefficiency \\
\hline
\cite{Ju2020Distributed,Khanapuri2023Learning,Xiao2021Secure}
& Platoon attacks
& Sensor tampering; comm. jamming
& Instability; collision risk \\
\hline
\cite{Zhang2020CANsec:,Li2021CAN}
& CAN bus hack
& Rogue commands
& Physical control takeover \\
\hline
\cite{Azmoodeh2018Detecting,Humayun2020Internet}
& Ransomware
& System lock
& Safety risk; operational halt \\
\bottomrule
\end{tabularx}
\end{table*}
\subsection{Threat Landscape}
\label{sec:threats}
Understanding the threat landscape is essential for identifying the limitations of conventional security models and motivating the adoption of emerging solutions. Figure~\ref{fig:IoVThreatModel} illustrates common attack vectors across the IoV stack, while Figure~\ref{fig:Incidents_by_Attack_vectors} shows that recent incidents most frequently target telematics units (43\%), infotainment systems (15\%), and public APIs (13\%). Table~\ref{tab:iov_threats} summarizes these threats and their primary points of entry. Attacks can be broadly categorized as follows:  
\textit{Network centric attacks}, such as Sybil and DoS, are commonly mitigated using distributed trust models and entropy based filtering~\cite{Li2021RTED-SD:,Sherazi2019DDoS,zhang2019lptd}.  
\textit{Identity based threats}, including GPS spoofing and certificate forgery, are addressed through Long Short-Term Memory (LSTM) based sensor fusion techniques~\cite{Dasgupta2021Sensor,Wazid2019AKM-IoV:}.  
\textit{Data integrity breaches} are countered by timestamp verification and encrypted cooperative perception~\cite{Katragadda2020Detecting,Song2020FBIA:}.  
\textit{Physical and software exploits}, such as CAN bus injection and unsecured APIs, require hardened gateways and encrypted internal communication~\cite{Zhang2020CANsec:,Ju2020Distributed,Chen2019A}.

\bolditalic{Implications:}~The threat landscape highlights the multidimensional nature of IoV security risks, which span from physical-layer intrusions to protocol level and application layer exploits. These observations underscore the limitations of traditional perimeter based defenses in coping with dynamic, large scale, and latency sensitive vehicular environments. The complexity and distribution of these threats necessitate layered mitigation strategies that combine local anomaly detection, secure communication protocols, and resilient network architectures. Accordingly, understanding the characteristics of these attack vectors is essential for assessing the effectiveness of existing security architectures and identifying areas that demand more adaptive and context aware solutions.

\begin{figure}[!t]
  \centering
  \begin{minipage}[t]{0.48\linewidth}
    \centering
    \includegraphics[width=\linewidth]{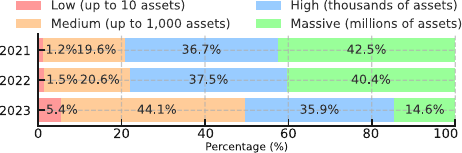}
    \caption{Impact of security incidents (year-wise \% of severity of incidents) in the IoV ecosystem.}
    \label{fig:iov_security_incidents}
  \end{minipage}\hfill
  \begin{minipage}[t]{0.48\linewidth}
    \centering
    \includegraphics[width=\linewidth]{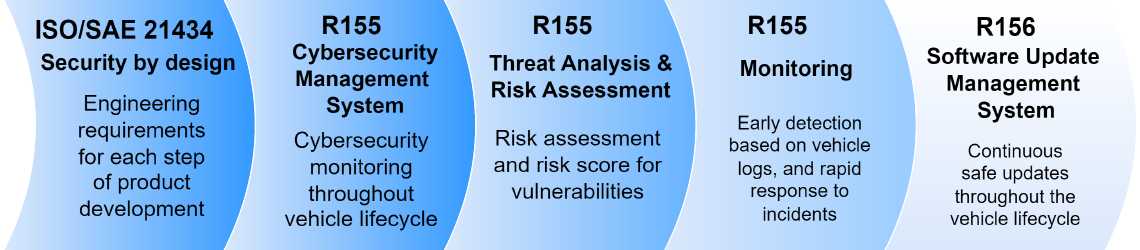}
    \caption{Cybersecurity standards and practices in IoV.}
    \label{fig:sae21434w29}
  \end{minipage}
\end{figure}
\subsection{Current State of IoV Security and Industry Trends}
\label{sec:state_and_trends}

Conventional IoV security frameworks rely on encryption, transport layer security, multi-factor authentication, and rule based IDS to establish baseline protections~\cite{Deebak2020A,Chen2019A,Rescorla2018The,Aman2021A,Sivanathan2020Managing}. While these mechanisms are effective against known threats, they face well documented limitations: deep packet inspection introduces latency overhead~\cite{Darwish2018Fog}, signature based detection fails to identify novel attacks~\cite{Niu2021Malware}, and centralized telemetry collection presents scalability challenges in dynamic vehicular environments~\cite{Ni2020Toward}.

\subsubsection{Modern countermeasures}
To address these limitations, industry and research efforts have increasingly focused on AI driven, data centric approaches. Early ML based IDS demonstrated improved adaptability to evolving threats~\cite{Bagga2021On}, while DL architectures such as convolutional neural networks (CNNs) and recurrent neural networks (RNNs) have achieved detection accuracies exceeding 99\%~\cite{Oseni2023An}. Hardware accelerators deployed at the network edge enable sub-10 ms inference latency~\cite{Hussain2019Machine}, facilitating near real time response. Federated Learning (FL) has been employed to reduce raw data transmission while preserving privacy~\cite{Mothukuri2021Federated-Learning-Based}. In parallel, advances in blockchain for integrity assurance and trust~\cite{Jiang2019Blockchain-Based,zhang2019tppr}, lightweight cryptographic protocols, and quantum resistant security primitives~\cite{Gupta2022Quantum-Defended} further enhance the robustness of IoV defenses.

Table~\ref{tab:comparison_iov_security} contrasts traditional and modern approaches, indicating a clear shift toward distributed analytics, continuous model adaptation, and decentralized enforcement. However, ongoing challenges include scaling edge based AI, complying with jurisdiction specific data protection laws, and ensuring auditability of security mechanisms~\cite{Chen2018Special}.

\begin{table*}[!b]
\centering
\caption{Comparison of traditional and current IoV security solutions.}
\label{tab:comparison_iov_security}
\fontsize{6pt}{7.2pt}\selectfont
\begin{tabularx}{\linewidth}{
  >{\hsize=0.5\hsize}Y
  >{\hsize=0.85\hsize}Y
  >{\hsize=1\hsize}Y
  >{\hsize=1.65\hsize}Y
}
\toprule
\textbf{Aspect} & \textbf{Traditional} & \textbf{Current} & \textbf{Advances \& Limitations} \\
\hline

Regulatory  
& Region-specific \textbf{\cite{Barati2020GDPR}}
& GDPR, NIST \textbf{\cite{Sullivan2019EU}}  
& Better global alignment; cross-border compliance challenges \\ \hline

Energy  
& High consumption \textbf{\cite{Fu2016A}}  
& AI/EC-optimized \textbf{\cite{Shuvo2023Efficient}}  
& Improved power management; DL remains energy-intensive \\ \hline

Data processing  
& Centralized, high latency \textbf{\cite{Darwish2018Fog}}  
& Edge-powered, low latency \textbf{\cite{Shi2020Communication-Efficient}}  
& Real-time responses; potential edge bottlenecks \\ \hline

Threat detection  
& Signature-based \textbf{\cite{Niu2021Malware}}  
& ML/DL-driven \textbf{\cite{Oseni2023An}}  
& Proactive analytics; high compute and false positives \\ \hline

Security protocols  
& Static rules \textbf{\cite{Alshamrani2019A}}  
& Adaptive models \textbf{\cite{Zhang2022Multiaccess}}  
& Responsive to new threats; complexity and compliance risks \\ \hline

Interoperability  
& Device-specific \textbf{\cite{Ni2020Toward}}  
& Standardized protocols \textbf{\cite{Dai2019Industrial}}  
& Broad compatibility; legacy integration issues \\ \hline

Scalability  
& Poor at scale \textbf{\cite{Wu2019Mobility}}  
& Modular AI/EC \textbf{\cite{Popoola2021Federated}}  
& Handles large data; demands robust infrastructure \\ \hline

System updates  
& Manual, infrequent \textbf{\cite{chen2019secure}}  
& Automated, real-time \textbf{\cite{Bhat2020Edge}}  
& Continuous adaptation; update rollout risks \\ 
\bottomrule
\end{tabularx}
\end{table*}

\subsubsection{Industry pulse}
Market trends reflect the increasing importance of connected vehicle security. The global connected car market is projected to grow from \$80.0B in 2023 to \$92.6B in 2024, driven largely by 5G infrastructure and AI integration~\cite{businessresearch2023}. Concurrently, security incidents are escalating: Figure~\ref{fig:iov_security_incidents} shows that API-related breaches have risen by 380\% since 2021~\cite{upstream2024,upstream2023}. Notable cases include remote control exploits affecting over 25 automotive OEMs and a \$2.25M Bitcoin extortion targeting NIO~\cite{nio2023,securitymag2023}.

\subsubsection{Standardization and regulation.}
To address these systemic risks, regulatory and standards bodies have introduced security guidelines. ISO/SAE 21434 and UNECE WP.29 emphasize cybersecurity by design principles, secure software updates, and incident response readiness (Figure~\ref{fig:sae21434w29})~\cite{Costantino2022In-Depth,unece2024,ISO_SAE_21434_2021}. Complementing these, the NIST Cybersecurity Framework encourages continuous monitoring, over the air (OTA) patching, and automated compliance verification~\cite{nist2023}. Overall, the current landscape reflects a transition from static, infrastructure centric defenses to dynamic, AI-augmented and standards-aligned approaches. The following sections examine the role of EC, ML, and DL in enabling this evolution, with attention to both implementation strategies and open research challenges.

\subsection{Landscape of EC, ML, and DL for IoV Security}\label{sec:landscape_iov_security}
The increasing complexity and scale of IoV environments have driven a shift toward architectures that integrate EC, ML, and DL as part of the security infrastructure. These technologies are now being incorporated across system layers to support low latency detection, adaptive threat response, and privacy aware processing.

\subsubsection{Edge Computing Developments}
MEC servers embedded at RSUs increasingly run containerized security services, including certificate validation, local filtering, and misbehavior logging~\cite{Grover2021Edge}. Offload schedulers utilize vehicle context, such as speed or signal quality, to decide whether to transmit raw data, features, or inference results~\cite{xu2019edge}. Certificateless cryptographic schemes reduce handshake latency~\cite{xie2022efficient}, while fog based clustering improves fault tolerance by replicating services across RSUs~\cite{nsouli2023reinforcement}. Open challenges include coordinating trust across heterogeneous infrastructure and securing edge hardware against side-channel threats.

\subsubsection{ML Approaches}
Supervised learning methods, including Random Forest and SVM, remain prevalent in IDS, achieving high accuracy on datasets like CIC-IDS-2018 and VeReMi~\cite{talpur2021machine,qayyum2020securing}. Unsupervised techniques offer potential for zero-day detection but suffer from high false positives in dense traffic~\cite{boualouache2023survey}. FL is used to limit raw data transmission while supporting distributed training~\cite{deng2020edge}, and graph-based models enable context aware threat scoring based on vehicle interactions~\cite{yuce2024misbehavior}. Key limitations include data imbalance, limited onboard compute, and susceptibility to adversarial inputs.

\subsubsection{DL Architectures}
DL models support a range of tasks across the IoV stack. CNNs and LSTMs have been applied to RF fingerprinting, LiDAR analysis, and CAN bus anomaly detection~\cite{almehdhar2024deep,Chen2019Deep}. Lightweight autoencoders and variational models have been explored for real time anomaly detection on constrained hardware~\cite{Dai2019Artificial}. Layer splitting techniques enable early stage processing on vehicles with later stage inference on edge servers, reducing onboard load while preserving low-latency response~\cite{yang2020offloading}. Persistent gaps remain in handling concept drift and ensuring robustness against adversarial examples.

\subsubsection{Integration Patterns}
Three implementation patterns are common: (1) edge-hosted inference, where RSUs perform DL based threat detection with sub-20 ms latency~\cite{Grover2021Edge}; (2) hierarchical learning, combining local filtering at RSUs with cloud level model aggregation~\cite{liu2023crs}; and (3) collaborative analytics, where blockchain records model hashes and verdicts to ensure auditability~\cite{wang2023blockchain}. However, these components are often deployed in isolation; a unified architecture that jointly optimizes model placement, data privacy, and detection accuracy remains to be fully realized.

\section{EC as an Enabler of IoV Security}
\label{sec:role-edge-computing}
The previous section outlined the architectural structure of IoV systems and highlighted the security threats that arise from centralized processing. In response to these limitations, EC has emerged as a foundational enabler of scalable, low latency, and privacy preserving security frameworks. By relocating computation closer to vehicles, via OBUs and RSUs, EC reduces decision latency, limits attack surfaces, and supports localized analytics. This section surveys the role of EC in IoV security, focusing on processing architectures, supported security functions, and outstanding research challenges.

\subsection{Edge-Centric Processing Architectures in IoV Security}\label{sec:ec_fundamentals}
\begin{wrapfigure}{r}{0.5\textwidth} 
    \vspace{-10pt} 
    \centering
    \includegraphics[width=0.48\textwidth]{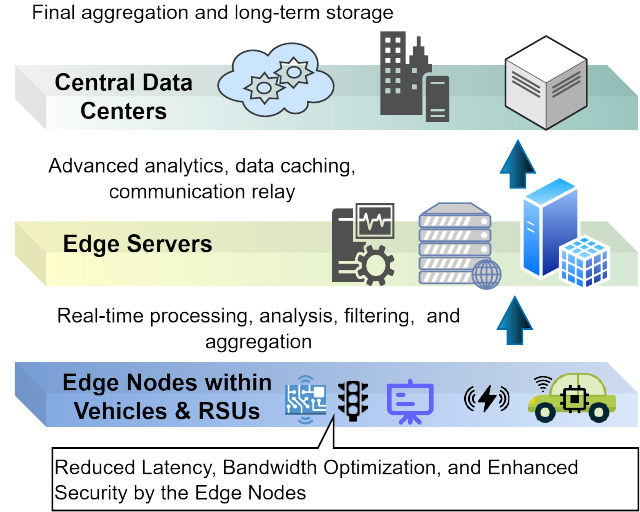}
    \caption{Generalized EC architecture for IoV.}
    \label{fig:ECArchitecture}
    \vspace{-10pt} 
\end{wrapfigure}

EC introduces a distributed processing paradigm within the IoV security architecture, enabling low latency threat detection, localized decision making, and reduced reliance on centralized cloud infrastructures. Figure~\ref{fig:ECArchitecture} illustrates a commonly adopted three tier architecture, comprising vehicle level micro edge nodes, RSUs, and cloud based services. This hierarchy reflects increasing processing capacity and decreasing responsiveness from edge to core. At the lowest tier, OBUs process high frequency sensor data streams such as CAN bus messages, LiDAR point clouds, and GPS telemetry. Lightweight models, typically 1D CNNs or statistical filters, are employed to discard malformed packets or detect basic anomalies in real time~\cite{Hou2020Reliable,mahadik2023edge}. These micro edge functions reduce the computational burden on RSUs while enabling initial threat filtering at the data source. The intermediate tier consists of RSUs equipped with MEC servers that support more computationally intensive tasks. Models such as GRU-based autoencoders or CNN-LSTM hybrids are deployed to identify stealthier threats, including distributed DDoS bursts, spoofed beacons, or protocol misuse~\cite{Shi2016Edge,Lin2019Computation}. RSUs may also perform data aggregation, rate limiting, and policy enforcement based on local context. At the top tier, cloud services handle global threat correlation, policy dissemination, and model refinement. Rather than transmitting raw sensor data, edge nodes forward meta-features or encrypted gradient updates, preserving privacy and reducing bandwidth consumption~\cite{Cui2020A,Xu2021Trust-Aware,Cui2021A}. FL paradigms are commonly used to support collaborative model improvement across distributed sites without centralized data collection~\cite{Yu2021EC-SAGINs:}.

\begin{figure*}[!t]
    \centering
    \includegraphics[width=0.8\textwidth]{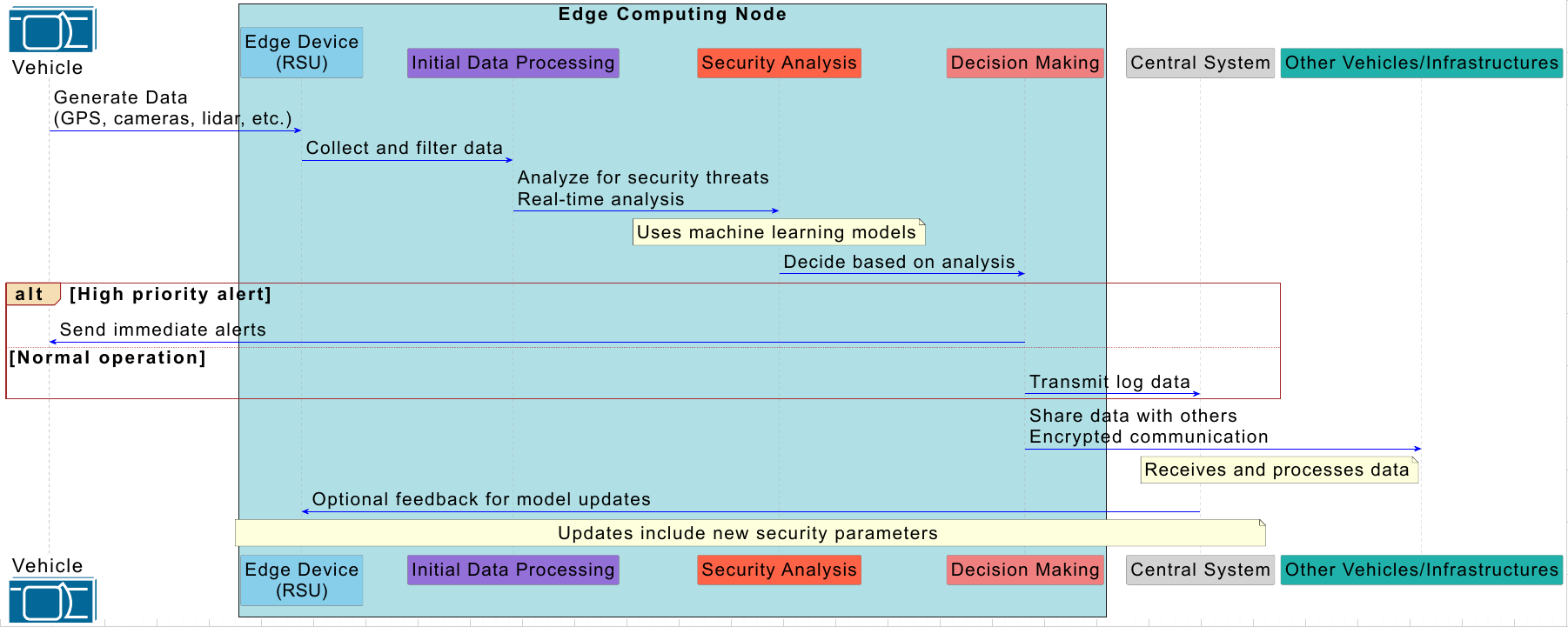}  
    \caption{EC in IoV security: Data flow \& real time processing between vehicles, edge nodes, \& central server.}
    \label{fig:DataFlowatECN}
\end{figure*}

Figure~\ref{fig:DataFlowatECN} illustrates the secure data flow in an EC-enabled IoV architecture. Vehicles generate sensor data that is transmitted to nearby RSUs for real-time analysis and initial threat detection. High-priority alerts trigger immediate broadcasts to surrounding vehicles to support cooperative awareness~\cite{wang2025reliability}. In cases of uncertainty, RSUs forward encrypted feature summaries to the central system for further evaluation. Model updates and refined security parameters are periodically sent back to edge nodes, enabling adaptive local detection. This tiered architecture balances responsiveness, scalability, and security while enabling fine-grained policy enforcement and coordinated defense.

\subsection{Security Functions Enabled by EC}\label{sec:ec_security_capabilities}
EC has been widely adopted in recent IoV security frameworks due to its ability to support localized decision making, reduce data exposure, and improve system responsiveness. This subsection surveys how existing research leverages EC to enable core security functions, grouped by capability domain.

\subsubsection{Reduced Exposure and Attack Surface}
Several studies emphasize EC’s role in minimizing the system’s exposure to external threats. Zhao et al. demonstrate that executing threat detection tasks at RSUs reduces the number of packets transmitted over public networks, thereby limiting opportunities for interception and MitM attacks~\cite{Zhao2018Deploying}. Quarantining compromised nodes at the edge, before their behavior influences global analytics, has also been proposed as a containment strategy.

\subsubsection{Data Protection and Confidentiality}
The work by Zhang et al. introduces LVPDA, a lightweight protocol tailored for resource constrained OBUs, enabling secure communication without significant memory overhead~\cite{Zhang2020LVPDA:}. In a similar direction, Fan et al. propose DR-BFT, a blockchain-based framework that enables tamper evident logging with minimal RSU storage requirements~\cite{Fan2021DR-BFT:}. Table~\ref{tab:EC_advantages_iov_solutions} summarizes several design approaches that preserve privacy at the edge. Techniques such as attribute based encryption and homomorphic aggregation are employed to ensure that only processed insights, not raw data, are shared with cloud infrastructure.

\subsubsection{Low Latency Threat Detection}
Researchers have explored deploying layered detection models to balance speed and accuracy. Cui et al. evaluate shallow models for anomaly detection at micro edge devices, enabling sub-millisecond inference for coarse threats~\cite{Cui2021Efficient}. In contrast, deeper models such as CNN-LSTM hybrids deployed at RSUs are used to detect stealthy or high-dimensional attacks. Table~\ref{tab:edge_security_features} outlines use cases where these models are particularly effective in time-critical scenarios. Work by Li et al. further integrates authentication mechanisms using aggregate signatures and decentralized token checks directly at the edge~\cite{Li2021Inspecting}.

\begin{table*}[!b]
\centering
\caption{Key advantages of EC in IoV.}
\label{tab:EC_advantages_iov_solutions}
\fontsize{6pt}{7.2pt}\selectfont
\begin{tabularx}{\linewidth}{%
  >{\hsize=0.6\hsize}Y
  >{\hsize=0.7\hsize}Y
  >{\hsize=1.5\hsize}Y
  >{\hsize=1.2\hsize}Y
}
\toprule
\textbf{Refs} & \textbf{Advantage} & \textbf{Description} & \textbf{Use Case} \\
\hline
\cite{Zhao2018Deploying,Ren2019Collaborative}
& Bandwidth opt.
& Reduces upstream data; eases network congestion
& Smart-city traffic mgmt \\ \hline

\cite{Cui2020A,Dai2020A}
& Enhanced privacy
& Local data processing; lowers breach risk
& AV location sharing \\ \hline

\cite{Hou2020Reliable,Abouaomar2021Resource}
& Latency cut
& Edge-side analytics; faster response
& Real-time obstacle avoidance \\ \bottomrule
\end{tabularx}
\end{table*}

\subsubsection{Operational Efficiency and Proactive Response}
Edge enabled architectures also support broader system level goals. Hou et al. and Abouaomar et al. show that local inference reduces communication overhead and radio congestion in dense vehicular environments~\cite{Hou2020Reliable,Abouaomar2021Resource}. Work by Cui et al. and Zhu et al. demonstrates that predictive models deployed at RSUs can preemptively detect intrusion and unsafe driving conditions, enabling real time actuation in V2X applications~\cite{Cui2020A,Bilal2024RobustIoV}. Overall, the literature confirms that EC strengthens both the defensive and operational dimensions of IoV security. By pushing intelligence closer to data sources, EC enables scalable and context aware mechanisms that align with the timing, privacy, and resource constraints of connected vehicle ecosystems.

\begin{table*}[!t]
\centering
\caption{Key security features enabled by EC in IoV and real-world applications.}
\label{tab:edge_security_features}
\fontsize{6pt}{7.2pt}\selectfont
\begin{tabularx}{\linewidth}{
  >{\hsize=1.0\hsize}Y
  >{\hsize=0.375\hsize}Y
  >{\hsize=0.375\hsize}Y
  >{\hsize=0.375\hsize}Y
  >{\hsize=0.375\hsize}Y
  >{\hsize=0.375\hsize}Y
  >{\hsize=0.375\hsize}Y
  >{\hsize=0.375\hsize}Y
  >{\hsize=0.375\hsize}Y
}
\toprule
\textbf{Ref} 
& \textbf{Latency↓} 
& \textbf{FW‐Upd} 
& \textbf{Encrypt} 
& \textbf{Detect} 
& \textbf{Auth} 
& \textbf{Access} 
& \textbf{Integrity} 
& \textbf{Anomaly} \\
\hline
\cite{Shuvo2023Efficient}
& \checkmark & $\times$ & $\times$ & $\times$ & $\times$ & $\times$ & $\times$ & $\times$ \\
\hline
\cite{Zhang2020Blockchain-based}
& $\times$ & \checkmark & $\times$ & $\times$ & $\times$ & $\times$ & $\times$ & $\times$ \\
\hline
\cite{Shi2020Communication-Efficient}
& $\times$ & $\times$ & \checkmark & $\times$ & $\times$ & $\times$ & $\times$ & $\times$ \\
\hline
\cite{Oseni2023An}
& $\times$ & $\times$ & $\times$ & \checkmark & $\times$ & $\times$ & $\times$ & $\times$ \\
\hline
\cite{Dai2019Industrial}
& $\times$ & $\times$ & $\times$ & $\times$ & \checkmark & $\times$ & $\times$ & $\times$ \\
\hline
\cite{Popoola2021Federated}
& $\times$ & $\times$ & $\times$ & $\times$ & $\times$ & \checkmark & $\times$ & $\times$ \\
\hline
\cite{chen2019secure}
& $\times$ & $\times$ & $\times$ & $\times$ & $\times$ & $\times$ & \checkmark & $\times$ \\
\hline
\cite{Bourechak2023At}
& $\times$ & $\times$ & $\times$ & $\times$ & $\times$ & $\times$ & $\times$ & \checkmark \\
\bottomrule
\end{tabularx}
\end{table*}

\subsection{Open Challenges and Future Directions in EC-Enabled IoV Security}\label{sec:ec_challenges_future}
EC has enabled advancements in IoV security, several technical and architectural challenges remain unresolved. This subsection surveys key limitations identified in the literature and highlights promising directions for future research, with emphasis on deployment scalability, trust management, and convergence with emerging technologies.

\subsubsection{Resource Allocation and Model Placement}
Efficient distribution of computation between vehicle, RSU, and cloud tiers remains a core challenge. Hou et al. and Zhao et al. discuss the difficulty of supporting GPU-class inference at scale across RSUs~\cite{Hou2020Reliable, Zhao2021Edge}. Existing work explores dynamic offloading strategies that consider local queue length, wireless channel state, and model complexity, but real-time guarantees and generalizability across scenarios remain open problems~\cite{Fan2021DR-BFT:,Grover2021Edge}.

\subsubsection{Trust Management and Security Foundations}
Establishing trust at the edge involves trade-offs between performance and cryptographic rigor~\cite{Xu2020Secure, Wang2020Topology}. Xu et al. and Wang et al. emphasize the need for tamper proof hardware, secure boot processes, and continuous attestation to prevent side channel attacks on exposed edge nodes~\cite{Grover2021Edge}. Grover et al. highlight the complexity of key revocation in local escrow schemes~\cite{fernandez2019pre}. Fernandez et al. propose integrating post quantum cryptography to future-proof V2X communication.

\subsubsection{Scalability and Heterogeneity}
EC deployments face management overhead due to heterogeneous hardware, varied software stacks, and patching complexity~\cite{Grover2021Edge}. Studies have proposed containerized, OTA update mechanisms to streamline DevSecOps for distributed edge environments, yet coordinating updates across thousands of mobile and static edge nodes introduces reliability and consistency concerns.

\subsubsection{Privacy Preserving Learning and Concept Drift}
Adaptive security frameworks based on FL must account for rapidly changing traffic conditions and adversarial patterns~\cite{deng2020edge}. Deng et al. propose scheduling update intervals to manage radio contention in vehicular platoons~\cite{Xu2021Trust-Aware}. Xu et al. examine trust-aware aggregation in the presence of compromised edge nodes. However, maintaining model accuracy under concept drift without compromising privacy remains an active area of research.

\subsubsection{Integration with Emerging Technologies}
Recent studies envision the convergence of EC with complementary technologies~\cite{Zhou2019Edge}. Zhou et al. investigate continual learning for predictive, rather than reactive, threat detection~\cite{Kong2022Edge}. Lightweight blockchains for telemetry integrity, quantum resistant cryptographic schemes, and deterministic networking over 5G/6G links are increasingly proposed as architectural extensions~\cite{fernandez2019pre, Liu2020Toward}. Edge-rendered augmented reality (AR) dashboards have also been explored for real-time threat visualization~\cite{Premsankar2018Edge}.

Table~\ref{tab:EC_technological_advancements_iov} summarizes these future facing developments. Integrating such technologies into a unified, resource aware edge intelligence stack that complies with both safety and regulatory requirements presents a key direction for the next generation of IoV systems.

\begin{table*}[!t]
\centering
\caption{Future EC technologies for IoV security: technology use (\checkmark/ $\times$) and their impact.}
\label{tab:EC_technological_advancements_iov}
\fontsize{6pt}{7.2pt}\selectfont
\begin{tabularx}{\linewidth}{%
  >{\hsize=0.80\hsize}Y  
  >{\hsize=0.35\hsize}Y  
  >{\hsize=0.35\hsize}Y  
  >{\hsize=0.35\hsize}Y  
  >{\hsize=0.35\hsize}Y  
  >{\hsize=0.35\hsize}Y  
  >{\hsize=0.35\hsize}Y  
  >{\hsize=0.35\hsize}Y  
  >{\hsize=0.75\hsize}Y  
}
\toprule
\textbf{Refs} & \textbf{AI} & \textbf{QC} & \textbf{BC} & \textbf{AR/VR} & \textbf{5G/6G} & \textbf{Pred Ana.} & \textbf{Auto} & \textbf{Impact} \\
\hline
\cite{Zhou2019Edge}                          & \checkmark & $\times$ & $\times$ & $\times$ & $\times$ & $\times$ & $\times$ & Threat Detection \\ \hline
\cite{fernandez2019pre}                      & $\times$   & \checkmark & $\times$ & $\times$ & $\times$ & $\times$ & $\times$ & Data Protection \\ \hline
\cite{Kong2022Edge,zhang2020bsfp,zhang2020prvb} & $\times$ & $\times$ & \checkmark & $\times$ & $\times$ & $\times$ & $\times$ & Integrity \\ \hline
\cite{Premsankar2018Edge}                    & $\times$ & $\times$ & $\times$ & \checkmark & $\times$ & $\times$ & $\times$ & Situational Awareness \\ \hline
\cite{Liu2020Toward,Pham2019A}               & $\times$ & $\times$ & $\times$ & $\times$ & \checkmark & $\times$ & $\times$ & Low Latency \\ \hline
\cite{Wang2020Deep} (Pred.\ Anal.)           & $\times$ & $\times$ & $\times$ & $\times$ & $\times$ & \checkmark & $\times$ & Proactive Defence \\ \hline
\cite{Wang2020Deep} (Auto Edge Mgmt)              & $\times$ & $\times$ & $\times$ & $\times$ & $\times$ & $\times$ & \checkmark & Continuous Integrity \\ 
\bottomrule
\end{tabularx}
\end{table*}

\section{ML Techniques in IoV Security}
\label{sec:ml-techniques}

ML is transforming threat detection and mitigation in IoV security. This section explores how ML enhances detection capabilities, integrates with IoV security frameworks, and offers improvements over traditional methods. We discuss specific strategies, challenges, and future directions for ML in IoV security.
\begin{table*}[!b]
\centering
\caption{ML paradigms for IoV security.}
\label{tab:ml_paradigms_iov}
\fontsize{6pt}{7.2pt}\selectfont
\begin{tabularx}{\linewidth}{%
  >{\hsize=0.5\hsize}Y
  >{\hsize=1.2\hsize}Y
  >{\hsize=1.2\hsize}Y
  >{\hsize=1.1\hsize}Y
}
\toprule
\textbf{Paradigm} & \textbf{Strengths} & \textbf{Limitations} & \textbf{Typical IoV Use} \\
\hline
Supervised
& \checkmark High accuracy on known patterns
& $\times $ Needs labeled data; weak on zero-day
& Malware / IDS (known attacks) \\ \hline
Unsupervised
& \checkmark Detects unknowns; unlabeled data
& $\times $ Many false positives; needs validation
& V2X anomaly spotting \\ \hline
Reinforcement
& \checkmark Adapts to dynamics; policy learning
& $\times $ Compute heavy; reward tuning
& Adaptive IDS; policy optimisation \\ 
\bottomrule
\end{tabularx}
\end{table*}

\subsection{Taxonomy of ML Paradigms in IoV Security}\label{sec:ml_fundamentals}
ML has become a foundational tool in IoV security, enabling adaptive, data driven defenses that address the limitations of static, rule based systems. This subsection categorizes the principal ML paradigms, supervised learning, unsupervised learning, and Reinforcement Learning (RL), and surveys their respective applications in IoV security contexts. Figure~\ref{fig:MLOverviewinIoV} illustrates the paradigm shift from static signature based detection to predictive analytics, while Table~\ref{tab:ml_paradigms_iov} aligns each learning category with representative security use cases.

\subsubsection{Supervised Learning}
Supervised learning methods remain the most widely adopted in IoV security research. Classifiers such as Random Forests, SVMs, and Gradient Boosting Machines can detect replayed beacons, forged packets, and malware with over 95\% accuracy on public datasets~\cite{Bilal2024RobustIoV}. Gao et al. and Li et al. further show that these models outperform traditional intrusion detection baselines in latency and precision~\cite{Gao2018A,Li2021Inspecting}. However, supervised models typically rely on labeled datasets and require retraining to remain effective against evolving threats, such as novel DoS variants.

\subsubsection{Unsupervised Learning}
Unsupervised methods are increasingly employed to detect previously unseen or zero-day threats. Clustering algorithms like k-Means and Self-Organizing Maps (SOM) and dimensionality reduction techniques like Principal Component Analysis (PCA) have been used to identify anomalous behavior in unlabeled vehicular traffic~\cite{Wickramasinghe2021Explainable}. These approaches are well suited for open V2V communication environments, where curated training labels are unavailable. However, they often suffer from high false positive rates and require human in the loop review or hybrid pipelines to improve decision quality.

\subsubsection{Reinforcement Learning}
RL offers a dynamic alternative to static policy enforcement by allowing agents to adaptively tune defense strategies based on real-time feedback. RL-based adaptation of firewall rules and quality of service settings has been explored in vehicular edge environments~\cite{Zeng2022APD:,Li2021Auditing}. While RL agents demonstrate high reactivity to probing attacks and stealthy behavior, their training and inference demands can strain resource-constrained RSUs. Ongoing research explores the use of hardware accelerators and model compression to mitigate this limitation.

\subsubsection{Hybrid and Layered Approaches}
Recent platforms integrate multiple ML paradigms to create layered security defenses. Supervised classifiers are deployed for well-characterized threats, unsupervised detectors monitor for emerging anomalies, and RL agents adapt firewall or routing policies in real time. This multi-paradigm strategy enables systems to balance accuracy, adaptability, and computational efficiency across varying IoV conditions, as surveyed in~\cite{Bilal2024RobustIoV,Gao2018A,Wickramasinghe2021Explainable,Zeng2022APD:}.

\begin{figure}[!t]
  \centering
  \begin{minipage}[t]{0.40\textwidth}
    \centering
    \includegraphics[width=\linewidth]{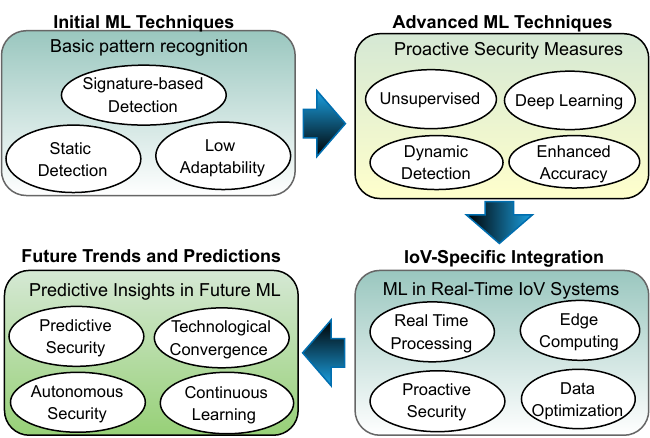}
    \caption{Evolution of ML in cybersecurity.}
    \label{fig:MLOverviewinIoV}
  \end{minipage}
  \hfill
  \begin{minipage}[t]{0.5\textwidth}
    \centering
    \includegraphics[height=0.7\linewidth,angle=90]{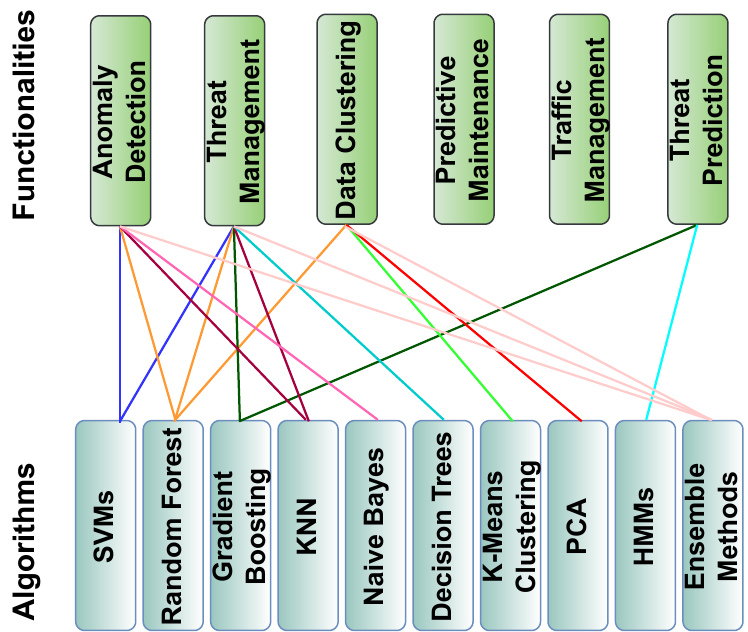}
    \caption{Relationship between ML algorithms \& their functionalities within IoV context.}
    \label{fig:alg_func_relationship}
  \end{minipage}
\end{figure}

\subsection{Representative Models and Use Cases}\label{sec:ml_use_cases}
This subsection surveys representative ML algorithms applied in IoV security and categorizes their deployment across common tasks such as intrusion detection, anomaly detection, behavior prediction, and flow classification. Figure~\ref{fig:alg_func_relationship} maps these models to their corresponding security functions, while Table~\ref{tab:ml_comparison} summarizes their reported performance on benchmark datasets.

\subsubsection{Classification and Filtering}
Supervised classifiers such as SVMs, Random Forests, and Gradient Boosting Machines are widely used for classifying benign versus malicious traffic. These models achieve over 95\% accuracy in detecting malware injections and replay attacks in V2X communications~\cite{Bilal2024RobustIoV,Gao2018A}. SVMs are particularly effective for high dimensional feature spaces due to their kernel-based separation, while ensemble methods like Random Forests improve robustness under class imbalance by aggregating weak learners. Lightweight classifiers such as $k$-Nearest Neighbors (k-NN), Naïve Bayes, and Decision Trees are still valuable for real-time filtering on constrained OBUs.

\subsubsection{Anomaly and Outlier Detection}
Unsupervised models play a critical role in detecting zero-day threats and protocol deviations. k-Means and PCA have been used to identify deviations in vehicular telemetry, while Hidden Markov Models (HMMs) have been applied to capture temporal anomalies in CAN bus messages and beacon sequences~\cite{Wickramasinghe2021Explainable}. These models enable continuous monitoring without requiring labeled data but typically involve trade-offs between sensitivity and false alarm rates.

\subsubsection{Intrusion Detection Enhancements}
Recent studies have advanced ML-based IDS using ensemble tuning, hybrid pipelines, and transfer learning. Intrusion detection accuracies exceeding 98\% have been reported after applying Bayesian optimization and semi-supervised reinforcement learning~\cite{Oseni2023An,Dong2021Network}. Further improvements in precision have been achieved by incorporating multi-stage classifiers, reputation scoring, and feature selection~\cite{Rashid2023An,Injadat2020Multi-Stage}. Transfer learning approaches address the scarcity of labeled data by fine-tuning pre-trained models on related vehicular datasets~\cite{Gyawali2020Machine}.

\subsubsection{Behavior Modeling and Prediction}
Predictive modeling in IoV is increasingly supported by sequence-aware models such as RNNs and LSTM networks. These architectures are well suited for modeling time series data from sensors or vehicle trajectories~\cite{Aslam2022Adaptive,Wang2020Deep}. Graph-based models are also emerging to represent dynamic vehicular relationships and infer cooperative threat behavior, though their deployment remains limited in practice.

The surveyed models collectively support a multi-layered defense strategy in IoV security systems. Their deployment depends on task specificity, available compute resources, and tolerance for false positives, with growing attention to hybrid and adaptive model pipelines.

\begin{table*}[!t]
\centering
\begin{threeparttable}
\caption{ML vs.\ traditional methods in cybersecurity ($\checkmark$ = better detection, $\downarrow$ = faster response).}
\label{tab:ml_comparison}
\fontsize{6pt}{7.2pt}\selectfont

\begin{tabularx}{\linewidth}{
  >{\hsize=0.6\hsize\raggedright\arraybackslash}X   
  >{\hsize=2\hsize\raggedright\arraybackslash}X   
  >{\hsize=0.7\hsize\centering\arraybackslash}X     
  >{\hsize=0.7\hsize\centering\arraybackslash}X     
}
\toprule
\textbf{Refs} & \textbf{ML Technique(s)} & \textbf{Detect \mbox{$\uparrow$}} & \textbf{Resp.\ \mbox{$\downarrow$}} \\
\hline
\cite{Li2021Transfer}                    & Transfer Learning                 & \mbox{$\checkmark$} & \mbox{$\checkmark$} \\ \hline
\cite{Oseni2023An}                       & DL + SHAP (Shapley Explainable)              & \mbox{$\checkmark$} & \mbox{$\checkmark$} \\ \hline
\cite{Injadat2020Multi-Stage}             & Bayesian \& Hyper-parameter optimization    & \mbox{$\checkmark$} & \mbox{$\checkmark$} \\ \hline
\cite{Rashid2023An}                      & Gradient Boosting; Logistic Reg.; Multi-Layer Perceptron; RF; SVM  & \mbox{$\checkmark$} & \mbox{$\checkmark$} \\ \hline
\cite{Gyawali2020Machine}                & SVM, Feed-forward NN               & \mbox{$\checkmark$} & \mbox{$\checkmark$} \\ \hline
\cite{Sharma2021A}                       & SVM, KNN, Naïve Bayes, RF                       & \mbox{$\checkmark$} & \mbox{$\checkmark$} \\ \hline
\cite{Abdelmoumin2021On}                 & Ensembles (bagging, boosting, stacking) & \mbox{$\checkmark$} & \mbox{$\checkmark$} \\ \hline
\cite{Dong2021Network}                   & Semi-supervised Deep RL                & \mbox{$\checkmark$} & \mbox{$\checkmark$} \\ 
\bottomrule
\end{tabularx}

\vspace{-2pt}
\begin{tablenotes}[flushleft]\scriptsize
\item $\uparrow$ Higher detection; $\downarrow$ Lower response time; $\checkmark$ = better than baseline.
\end{tablenotes} 
\end{threeparttable}
\end{table*}

\subsection{Architectural Integration of ML in IoV Security Frameworks}
\label{sec:ml_integration}

Effective deployment of ML in IoV security requires architectural integration across multiple tiers, OBUs, RSUs, edge gateways, and cloud infrastructure. Figure~\ref{fig:MLIntegrationinIoV} illustrates the common multi-tier pattern observed in recent studies. At the embedded layer, lightweight classifiers are deployed on OBUs and RSUs to support real-time anomaly detection based on control messages and sensor data. The work in Ullah et al. shows that such models can detect anomalies in CAN bus signals within milliseconds, allowing fail-safe triggers at the point of data generation~\cite{Ullah2022HDL-IDS:}.

At the network tier, edge nodes aggregate traffic across vehicles, enabling correlation of flow-level behavior. Aslam et al. demonstrate that clustering and supervised classification at this tier can identify coordinated scan patterns and distributed attacks that would be invisible to individual vehicles~\cite{Aslam2022Adaptive}. Beyond localized inference, the security operations tier supports centralized orchestration of alerts and automated responses. Oseni et al. and Li et al. describe ML-enhanced SOC platforms that incorporate ensemble classifiers, reputation-based threat scoring, and mitigation playbooks~\cite{Oseni2023An,Li2019System}. These systems bridge multiple tiers and ensure consistent policy enforcement across the IoV network. At the backend, cloud services manage global threat intelligence and retrain detection models using broader datasets. As demonstrated in Alsharif et al. and Javaid et al., deep models trained in the cloud can be distilled and pushed to the edge in compressed formats, supporting deployment on resource-constrained OBUs~\cite{Alsharif2021Study,Javaid2021A}. FL architectures further reduce data exfiltration by enabling local training and global model aggregation without exposing raw telemetry.

This layered integration facilitates a balance between responsiveness and depth. While embedded and edge models offer low-latency threat detection, cloud-tier analytics contribute broader visibility and continuous model evolution. As IoV networks scale, optimizing the distribution of ML capabilities across these layers will be essential for ensuring resilience and compliance with latency and privacy constraints.

\subsection{Recent Advancements and Novel ML Techniques for IoV Security}
\label{sec:ml_future}

Recent research in IoV security has explored several machine learning advancements aimed at enhancing adaptability, robustness, and scalability. These include \textit{Federated Learning}, \textit{adversarial training}, \textit{graph-based models}, \textit{self-supervised learning}, and emerging directions in \textit{quantum-assisted inference}.

The authors in Yamany et al. and Cao et al. propose \textit{Federated learning} to enable collaborative model training across OBUs and RSUs without centralizing raw data~\cite{Yamany2023OQFL:,Cao2022PerFED-GAN:}. These FL variants reduce bandwidth consumption and mitigate privacy risks, especially in scenarios involving sensitive GPS and sensor data. \textit{Adversarial robustness} is addressed through GAN-augmented pipelines, as shown in the work by Bourechak et al. and Zeng et al., where synthetic adversarial inputs are used to improve IDS resistance against evasion and poisoning attacks~\cite{Bourechak2023At,Zeng2022APD:}. 

To improve relational context in IoV security, researchers have introduced \textit{graph neural networks} that model vehicle-to-vehicle interactions and detect collaborative threat patterns~\cite{Magar2022Crystal,Gong2022Self-Paced}. Similarly, the use of \textit{sequence-aware models}, including RNNs and LSTMs, has enabled predictive threat detection based on historical vehicular telemetry~\cite{Wang2020Deep}. The work in Gong et al. further explores \textit{self-supervised learning} (SSL) to exploit unlabeled data by leveraging pretext tasks for representation learning~\cite{Gong2022Self-Paced}. SSL is particularly effective in dynamic traffic environments where labeled anomalies are rare or delayed. 

Lastly, \textit{quantum-assisted learning} is emerging as a long-term direction. Huang et al. highlight the potential of quantum ML to accelerate cryptographic analysis and anomaly inference, though practical deployment remains speculative~\cite{Huang2022Quantum}.

These advancements collectively represent a shift from reactive security to proactive and context-aware defense. Their integration into future IoV architectures will be critical for ensuring both fleet-wide protection and compliance with privacy-preserving constraints.

\subsection{Open Challenges and Research Directions}
\label{sec:ml_challenges}

Despite the progress in ML-driven IoV security, several persistent challenges hinder practical deployment and long term effectiveness. These challenges fall under four thematic areas: \textit{data availability}, \textit{model adaptivity}, \textit{system scalability}, and the path toward \textit{collaborative, self-evolving defenses}.

The authors in highlight the issue of \textit{data availability and quality}, where privacy constraints, inconsistent formats, and the high cost of annotation limit the creation of large-scale IoV datasets~\cite{Shen2019Privacy-Preserving,Li2019System}. Modalities such as LiDAR, CAN, and GPS require specialized preprocessing and present interoperability issues across platforms. Secure synthetic data generation and simulation-based augmentation are increasingly explored to address rare event coverage and data scarcity. \textit{Model drift and adaptivity} remain active concerns due to the non-stationary nature of IoV environments. Static ML models degrade quickly as attacker behaviors, network conditions, and traffic patterns evolve. Incremental learning and online RL approaches offer promising directions, but they remain limited by the compute and memory constraints of edge devices~\cite{Zeng2022APD:,Shen2019Privacy-Preserving}.

From a systems perspective, \textit{scalability and deployment complexity} challenge large-scale rollout. As noted by Grover et al. and Zhao et al., efficient orchestration of pruned models, over-the-air updates, and cross-node consistency in dynamic vehicular environments is difficult to maintain under mobility and bandwidth constraints~\cite{Grover2021Edge,Zhao2021Edge}. Finally, the vision of \textit{collaborative and self-evolving security} is gaining traction. FL, adversarial training, and self-supervised pipelines offer adaptive, distributed intelligence, yet require further evaluation under adversarial scenarios, real time response constraints, and privacy audits before full-scale deployment~\cite{Cao2022PerFED-GAN:,Yamany2023OQFL:,Bourechak2023At,Gong2022Self-Paced}. Addressing these challenges is central to building robust, privacy-aware, and continually evolving ML frameworks that can operate effectively at fleet scale in future IoV ecosystems.

\begin{figure}[!t]
  \centering
  \begin{minipage}[t]{0.48\textwidth}
    \centering
    \includegraphics[width=\linewidth]{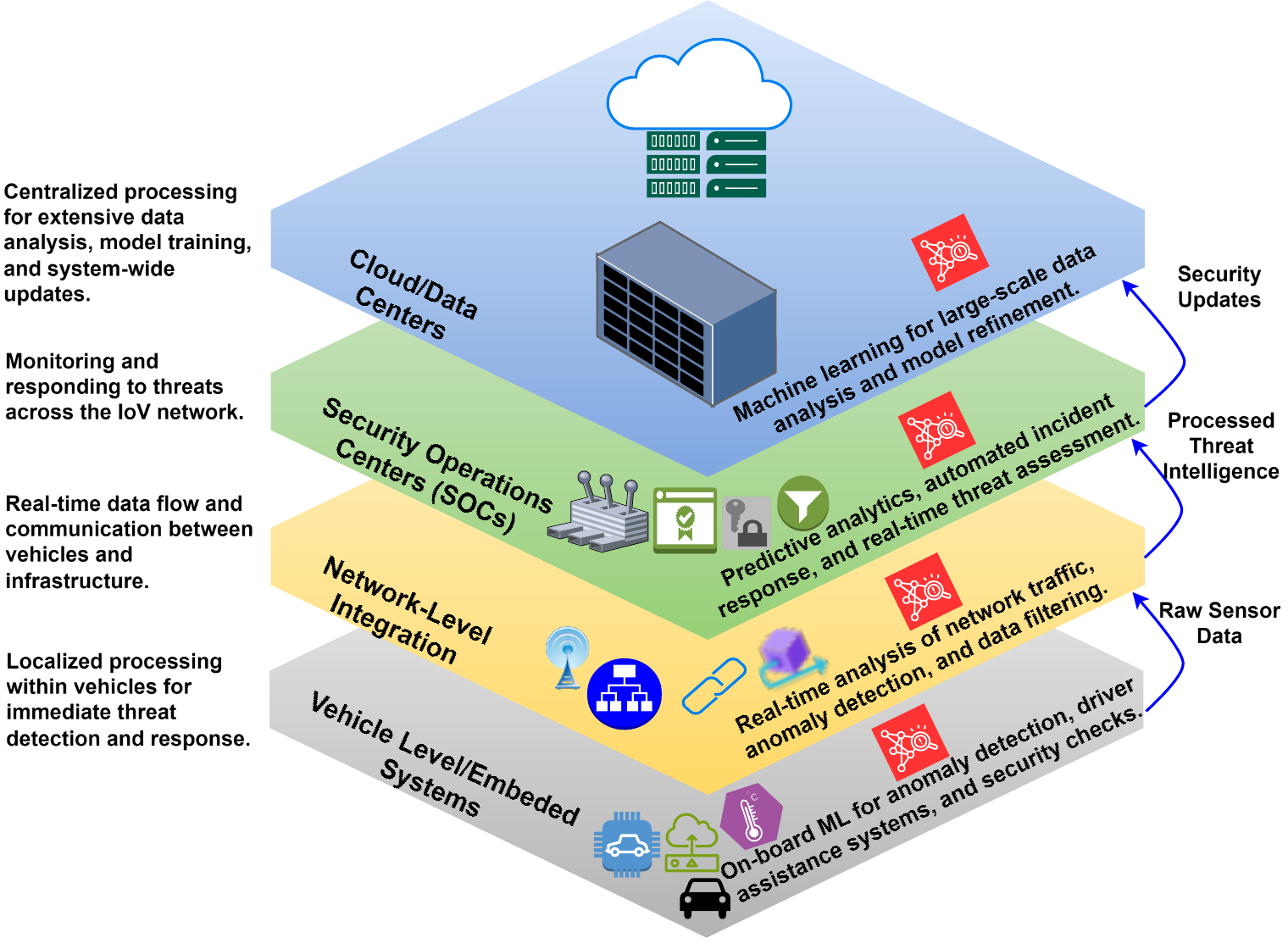}
    \caption{High-level architecture of ML integration within IoV security frameworks.}
    \label{fig:MLIntegrationinIoV}
  \end{minipage}
  \hfill
  \begin{minipage}[t]{0.48\textwidth}
    \centering
    \includegraphics[width=\linewidth]{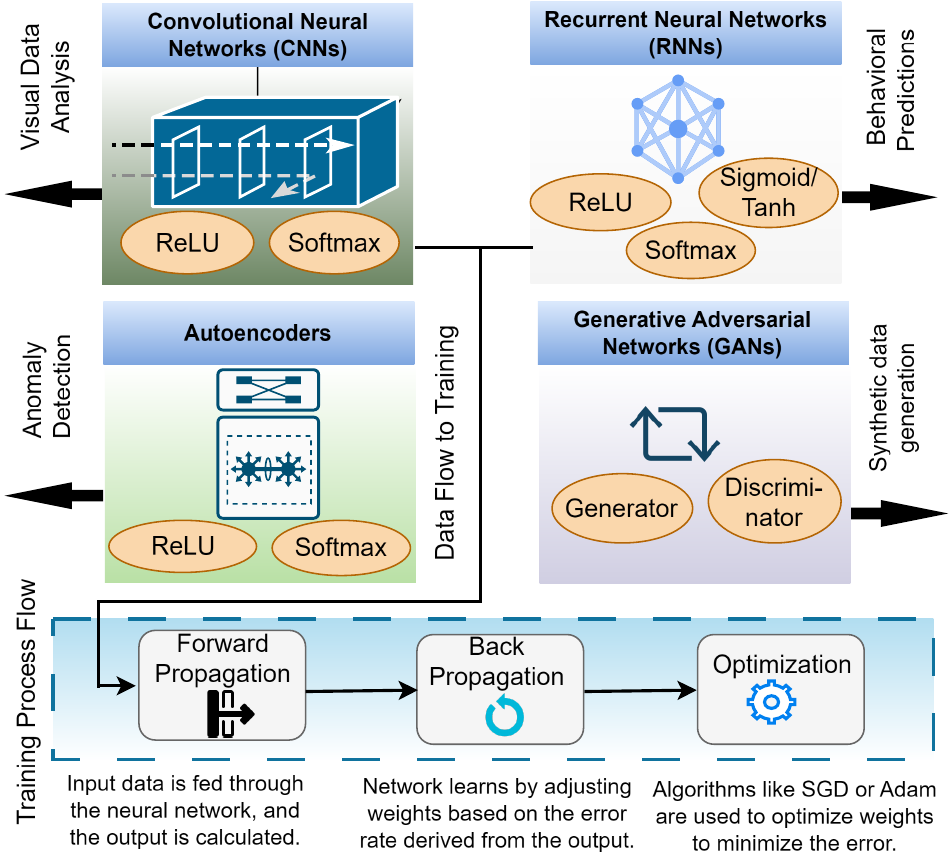}
    \caption{Core deep-learning architectures for IoV security: key roles and applications.}
    \label{fig:DLinIoV}
  \end{minipage}
\end{figure}

\begin{table*}[!b]
\centering
\caption{DL architectures in IoV: Advantages, Limitations, and Use Cases.}
\label{tab:deep_learning_architectures_iov}
\fontsize{6pt}{7.2pt}\selectfont
\begin{tabularx}{\linewidth}{%
  >{\hsize=0.5\hsize\raggedright\arraybackslash}X
  >{\hsize=1.2\hsize\raggedright\arraybackslash}X
  >{\hsize=1.2\hsize\raggedright\arraybackslash}X
  >{\hsize=1.1\hsize\raggedright\arraybackslash}X
}
\toprule
\textbf{Architecture} & \textbf{Advantage} & \textbf{Limitation} & \textbf{Use Case} \\
\hline
CNN  
& High image accuracy  
& High compute/memory  
& Traffic monitoring, obstacle detection \\ \hline

RNN  
& Temporal modeling  
& Vanishing gradients  
& Trajectory forecasting, flow analysis \\ \hline

AE  
& Anomaly detection  
& Noise sensitivity  
& IDS anomaly spotting \\ \hline

GAN  
& Data augmentation  
& Training instability  
& Synthetic attack data generation \\ 
\bottomrule
\end{tabularx}
\end{table*}

\section{Deep Learning for IoV Security: Architectures, Applications, and Challenges}\label{sec:dl-iov-security}
DL pushes IoV security beyond feature engineered ML by extracting multi scale patterns from images, signals, and traffic flows in real time.  Its layered representations unlock fine grained object detection for automated driving, sequence modelling for behaviour prediction, and unsupervised anomaly discovery in high velocity data streams.

\subsection{Deep Learning Architectures for IoV Security}
\label{sec:dl_fundamentals}

DL architectures form the backbone of many advanced IoV security systems by enabling direct learning from high-dimensional, multimodal inputs such as images, time series, and sensor telemetry. Table~\ref{tab:deep_learning_architectures_iov} summarizes key DL models and their applications in vehicular networks.

\textit{Convolutional neural networks} are extensively applied to spatially structured data including visual frames, RF signals, and radar inputs. Studies such as those by Nie et al. and Luo et al. demonstrate their utility in intrusion detection, spoofing classification, and real-time license plate recognition~\cite{Nie2020Data-Driven,luo2025anomaly}. While lightweight CNN variants are deployable at the edge, more complex models often require offloading to cloud or GPU-enabled RSUs. \textit{Recurrent neural networks} and \textit{LSTMs} are employed for sequence modeling, enabling temporal anomaly detection in CAN bus traffic and V2X communications. Grover et al. show that gated architectures preserve long-term dependencies and mitigate gradient vanishing, supporting detection of behavior-based threats~\cite{Grover2021Edge}. \textit{Autoencoders (AEs)} offer unsupervised detection by reconstructing inputs and flagging anomalies based on reconstruction loss. Ashraf et al. apply AEs to detect malformed and malicious traffic in vehicular networks, though their sensitivity to packet loss and noise remains a constraint~\cite{Ashraf2020Novel}. \textit{Transformer-based models} and hybrid CNN-Transformer designs have emerged to capture both local and global features in traffic logs. For instance, Zhu et al. propose FC-Trans, which improves rare-event detection through attention-based modeling~\cite{zhu2025fc}. These architectures are increasingly used for multimodal fusion and long-range dependency handling. \textit{Generative adversarial networks} are utilized to simulate adversarial behaviors and enrich training data for rare attacks. Xu et al. demonstrate GAN-based trace generation that improves generalization, though these models require tuning to avoid instability and mode collapse~\cite{Xu2019GANobfuscator:}.

Figure~\ref{fig:DLinIoV} illustrates a typical DL training pipeline, including forward and backward propagation, loss optimization via SGD or Adam, and regularization strategies such as dropout, batch normalization, and $L_2$ norms~\cite{LeCun2015Deep}. While these architectures significantly improve detection performance, achieving 10-20 ms inference on constrained hardware remains a core deployment challenge in real-world IoV environments.

\subsection{Applications of DL in Threat Detection and Behavior Analysis}\label{sec:dl_applications}
DL models support a range of IoV security functions, from intrusion detection to behavioral anomaly analysis. This subsection surveys key application domains where DL enhances real-time threat response, pattern recognition, and situational awareness in vehicular networks. Table~\ref{tab:deep_learning_architectures_iov} summarizes model types and their respective use cases.

\subsubsection{Intrusion and Threat Detection}
DL models are deployed for detecting intrusions, spoofing, malware propagation, and unauthorized access in IoV environments. CNNs have been shown to process image and video-based inputs from cameras and radar sensors for security monitoring and license plate verification~\cite{Nie2020Data-Driven}. Luo et al. propose lightweight CNN variants that maintain accuracy while operating under edge compute constraints~\cite{luo2025anomaly}. RNNs and LSTMs are used to monitor telemetry logs and communication sequences to flag temporal anomalies and potential protocol violations~\cite{Grover2021Edge}. Autoencoders have been employed for anomaly detection in network traffic, identifying deviations suggestive of DDoS attacks or malware injections~\cite{Ashraf2020Novel}. GANs, as demonstrated by Xu et al., are used to generate synthetic attack traces, improving model robustness against rare and evolving threats~\cite{Xu2019GANobfuscator:}.

\subsubsection{Scalability and Real Time Performance}
IoV networks generate high velocity data from sensors, cameras, and communication interfaces. CNNs’ parallelism enables scalable threat detection across multiple input streams~\cite{Qi2020Scalable}. Transfer learning has been used to fine-tune pre-trained models for domain-specific security tasks, reducing the need for large training datasets~\cite{Ning2021Intelligent}. These approaches enable DL systems to meet real-time constraints while maintaining high detection performance across heterogeneous vehicle fleets.

\subsubsection{Behavioral Modeling and Predictive Analytics}
DL models also support predictive security by modeling driving behavior, traffic flow, and communication patterns. Wang et al. and Ning et al. employ time series models to identify abnormal driving behaviors or communication outliers that could indicate hijacking or compromised control units~\cite{Wang2022Deep,Ning2019Deep}. DL has also been integrated into autonomous intrusion response systems such as REACT, which trigger local countermeasures in response to detected threats~\cite{hamad2024react}. These capabilities reduce dependence on centralized security operations and improve overall system responsiveness.

\subsubsection{Reducing False Positives and Alert Fatigue}
False alarms remain a significant challenge in large-scale IoV deployments. Studies have applied DL models such as autoencoders and GAN-based filters to refine anomaly scores and reduce redundant alerts~\cite{Ashraf2020Novel}. These methods improve signal to noise ratios for SOCs, enabling more targeted interventions and lowering the burden on human analysts.

\subsubsection{Hardware Acceleration for Edge DL}
To support real time inference, recent deployments incorporate DL-specific accelerators, such as GPUs and NPUs, into OBUs and RSUs. Wang et al. report significant gains in inference latency and energy efficiency using neural hardware~\cite{Wang2019A}. These advances enable DL models to operate at line speed, detecting threats without excessive computational overhead.

Together, these applications illustrate the breadth of DL's utility in securing IoV systems. From data rich perception tasks to lightweight behavioral inference, DL provides the adaptability and scalability required for modern vehicular cybersecurity.

\begin{table*}[!t]
\centering
\caption{IoV security protocols: evolution from traditional to adapted use and core impact.}
\label{tab:evolution_iov_security_protocols}
\fontsize{6pt}{7.2pt}\selectfont
\begin{tabularx}{\linewidth}{%
  >{\hsize=0.3\hsize\raggedright\arraybackslash}X
  >{\hsize=0.7\hsize\raggedright\arraybackslash}X
  >{\hsize=1.7\hsize\raggedright\arraybackslash}X
  >{\hsize=1.3\hsize\raggedright\arraybackslash}X
}
\toprule
\textbf{Refs} & \textbf{Protocol} & \textbf{Traditional Use → IoV Adaptation} & \textbf{Impact} \\
\hline
\cite{Chen2019A}  
& Digital Cert.  
& CA-issued web auth → Veh-infra identity vetting  
& Prevents spoofing; boosts trust \\ \hline

\cite{Zhang2020Blockchain-based}  
& Asym.\ Encryption  
& Email security → Secure V2V handshakes  
& Strong auth; ↑ compute cost \\ \hline

\cite{Rescorla2018The}  
& TLS  
& Web channel protection → V2V/V2I data integrity  
& Stops eavesdropping; ensures confidentiality \\ \hline

\cite{Deebak2020A}  
& Symm.\ Encryption  
& Pre-shared key streams → Real-time vehicle data encryption  
& Fast encrypt; key distrib.\ challenge \\ \hline

\cite{Alladi2021Artificial}  
& Firewall \& IDS  
& Network filtering → DPI-aware IoV traffic inspection  
& Blocks advanced threats; needs tuning \\ \hline

\cite{Jiang2019Blockchain-Based}  
& Blockchain  
& Financial ledger → Tamper-proof IoV logs \& updates  
& Immutable records; trust enhancement \\ \hline

\cite{Gupta2022Quantum-Defended}  
& Quantum Crypto  
& Future-proof comms → Quantum-safe key distribution  
& Unbreakable encryption; emerging tech \\ 
\bottomrule
\end{tabularx}
\end{table*}

\subsection{Trends in Lightweight and Privacy-Preserving DL}\label{sec:dl_trends}
As IoV systems grow in scale and complexity, emerging DL techniques are increasingly optimized for constrained edge environments and privacy-sensitive deployments. This subsection surveys the evolving landscape of lightweight DL architectures, privacy preserving training strategies, and protocol-level innovations that support secure and scalable inference in vehicular networks.

\subsubsection{Model Compression and Edge Optimization}
Recent work has emphasized reducing the size and computational load of DL models without sacrificing performance. Compression methods such as pruning, quantization, and knowledge distillation are widely used to meet the latency and energy constraints of OBUs and RSUs~\cite{Tung2020Deep,Prakash2022IoT}. Bhatia et al. demonstrate the feasibility of deploying CNNs on resource-constrained devices using depthwise separable convolutions and activation sparsity~\cite{Bhatia2022Improved}, while Luo et al. show that spectral residual filtering can enable anomaly detection at millisecond latency~\cite{luo2025anomaly}.

\subsubsection{Federated and Collaborative Learning}
To address privacy concerns and reduce transmission overhead, FL has been adopted in several IoV contexts. Architectures such as FED-IoV and VAN-FED-IDS distribute model training across RSUs and vehicles, enabling gradient aggregation without raw data sharing~\cite{Lu2020Blockchain,Lim2020Towards}. These systems enhance both privacy and scalability, particularly in compliance with regional data sovereignty regulations. Table~\ref{tab:evolution_iov_security_protocols} summarizes their protocol-level contributions.

\subsubsection{Online and Continual Learning}
To handle concept drift and the dynamic nature of IoV traffic, DL models must continuously adapt to new threat patterns. Lim et al. explore online learning frameworks that update model weights incrementally at the edge, reducing the delay between data collection and defensive response~\cite{Lim2020Towards}. Such systems require mechanisms for version control, automated validation, and rollback to ensure reliability.

\subsubsection{Security Protocol Innovations}
In parallel, DL-based security architectures are increasingly supported by advancements in underlying protocols. Blockchain-anchored logs provide immutable audit trails for model outputs~\cite{Zhu2020Two}, while quantum-resistant cryptographic schemes and secure firmware delivery protocols have been proposed to close the loop between detection, response, and recovery~\cite{Rescorla2018The,Deebak2020A,Zhang2020Blockchain-based}. These features are essential for building resilient, self-defending IoV systems that operate autonomously at scale.

Together, these trends signal a shift toward privacy-aware, low-latency, and decentralized DL pipelines that align with the real-world deployment constraints of vehicular environments. Continued progress in these directions will be critical for scaling AI-enabled security across diverse and mobile infrastructures.

\begin{table*}[!b]
\centering
\caption{Challenges, Impacts, and Solutions for IoV security systems using DL.}
\label{tab:iov_challenges_solutions}
\fontsize{6pt}{7.2pt}\selectfont
\begin{tabularx}{\linewidth}{%
  >{\hsize=0.4\hsize\raggedright\arraybackslash}X
  >{\hsize=1.0\hsize\raggedright\arraybackslash}X
  >{\hsize=1.1\hsize\raggedright\arraybackslash}X
  >{\hsize=1.5\hsize\raggedright\arraybackslash}X
}
\toprule
\textbf{Refs} & \textbf{Challenge} & \textbf{Solution} & \textbf{Notes} \\
\hline
\cite{Wang2019A}  
& $\uparrow$ Latency  
& HW/software pipeline optim.  
& Safety‐critical; infra compatibility \\ \hline

\cite{Chen2020Deep,Liang2021Pruning}  
& $\uparrow$ Compute  
& Model pruning \& quantization  
& Trade‐off: accuracy vs.\ size \\ \hline

\cite{Tung2020Deep}  
& Edge device constraints  
& Edge‐optimized AI accelerators  
& Requires edge‐specific hardware \\ \hline

\cite{Prakash2022IoT}  
& $\uparrow$ Energy use  
& Low‐power model design  
& Battery‐sensitive; lightweight arch. \\ \hline

\cite{Lu2020Blockchain}  
& $\uparrow$ Data needs  
& FL  
& Privacy‐preserving; comms overhead \\ \hline

\cite{Lu2020Blockchain}  
& Scalability issues  
& FL  
& Sync protocols for consistency \\ \hline

\cite{Zhao2020Local}  
& Privacy \& bias  
& Differential privacy \& encryption  
& Compliance critical; bias mitigation \\ \hline

\cite{Saputra2021Dynamic}  
& Continuous updates  
& Modular update frameworks  
& Downtime minimization; robust testing \\ \hline

\cite{Lim2020Towards}  
& Model drift  
& Online monitoring \& retraining  
& Automated anomaly detection \\ 
\bottomrule
\end{tabularx}
\end{table*}

\subsection{Deployment Challenges and Research Opportunities in DL-Based IoV Security}\label{sec:dl_deployment_challenges}
The integration of deep learning into IoV security frameworks must reconcile architectural, computational, and regulatory constraints to enable scalable, real-time protection. This subsection surveys current deployment strategies, system-level bottlenecks, and emerging research directions. Table~\ref{tab:iov_challenges_solutions} summarizes the key limitations and their corresponding mitigation approaches.

One critical concern is \emph{computational efficiency}. DL models such as CNNs and RNNs often exceed the processing and power budgets of OBUs, limiting their viability for latency-sensitive applications. The authors in Chen et al. and Prakash et al. show that uncompressed models compromise energy efficiency and responsiveness, i.e., delays beyond 100 ms can undermine real-time decision-making~\cite{Chen2020Deep,Prakash2022IoT,Wang2019A}. To address this, several works explore model optimization via quantization, pruning, and knowledge distillation~\cite{Tung2020Deep,Liang2021Pruning}, while Luo et al. propose architectural improvements such as depthwise convolutions and spectral residual filters for lightweight anomaly detection~\cite{luo2025anomaly}. Inference offloading to RSUs, as described by Grover et al. and Ning et al., reduces latency and bandwidth pressure, enabling sub-second threat responses~\cite{Grover2021Edge,Ning2019Deep}.

\emph{Data privacy and adaptability} pose additional barriers. Robust DL training requires large, diverse datasets, yet privacy regulations and proprietary restrictions often impede data centralization~\cite{Zhao2020Local}. Moreover, static models quickly degrade under evolving attack patterns and dynamic traffic flows. FL protocols such as FED-IoV and VAN-FED-IDS offer privacy-preserving alternatives by training local models and aggregating encrypted gradients~\cite{Lu2020Blockchain,Lim2020Towards}, reducing the risk of centralized breaches and supporting continuous learning. Table~\ref{tab:evolution_iov_security_protocols} outlines these advancements. However, concept drift and adversarial poisoning remain threats to model integrity across update cycles.

Scalability across heterogeneous devices introduces additional complexity. IoV environments are defined by intermittent connectivity, mobility, and diverse hardware capabilities. Bhatia et al. and Wang et al. highlight the role of hardware accelerators (e.g., NPUs, GPUs, ASICs) in restoring real-time inference on constrained platforms~\cite{Bhatia2022Improved,Wang2019A}. Still, as observed by Sonmez et al., model placement, scheduling, and policy adaptation must be co-designed with communication protocols to ensure performance stability under scale~\cite{Sonmez2020Machine}.

Looking ahead, DL-enabled IoV systems must evolve toward \emph{self-adaptive, trustworthy architectures}. Research by Zhu et al. suggests that blockchain-based audit trails, adversarial training, and continuous FL can support autonomous, privacy-compliant learning at scale~\cite{Zhu2020Two}. Yet, achieving secure, explainable, and regulatory-compliant DL integration across fleets remains an open challenge. Reliable over-the-air updates, tamper-evident model validation, and real-time resilience testing will be critical for production deployment. These challenges reflect the multidimensional nature of DL deployment in IoV; balancing responsiveness, scalability, energy efficiency, and privacy protection under real world constraints.

\begin{figure*}[!t]
    \centering
    \includegraphics[width=0.8\textwidth]{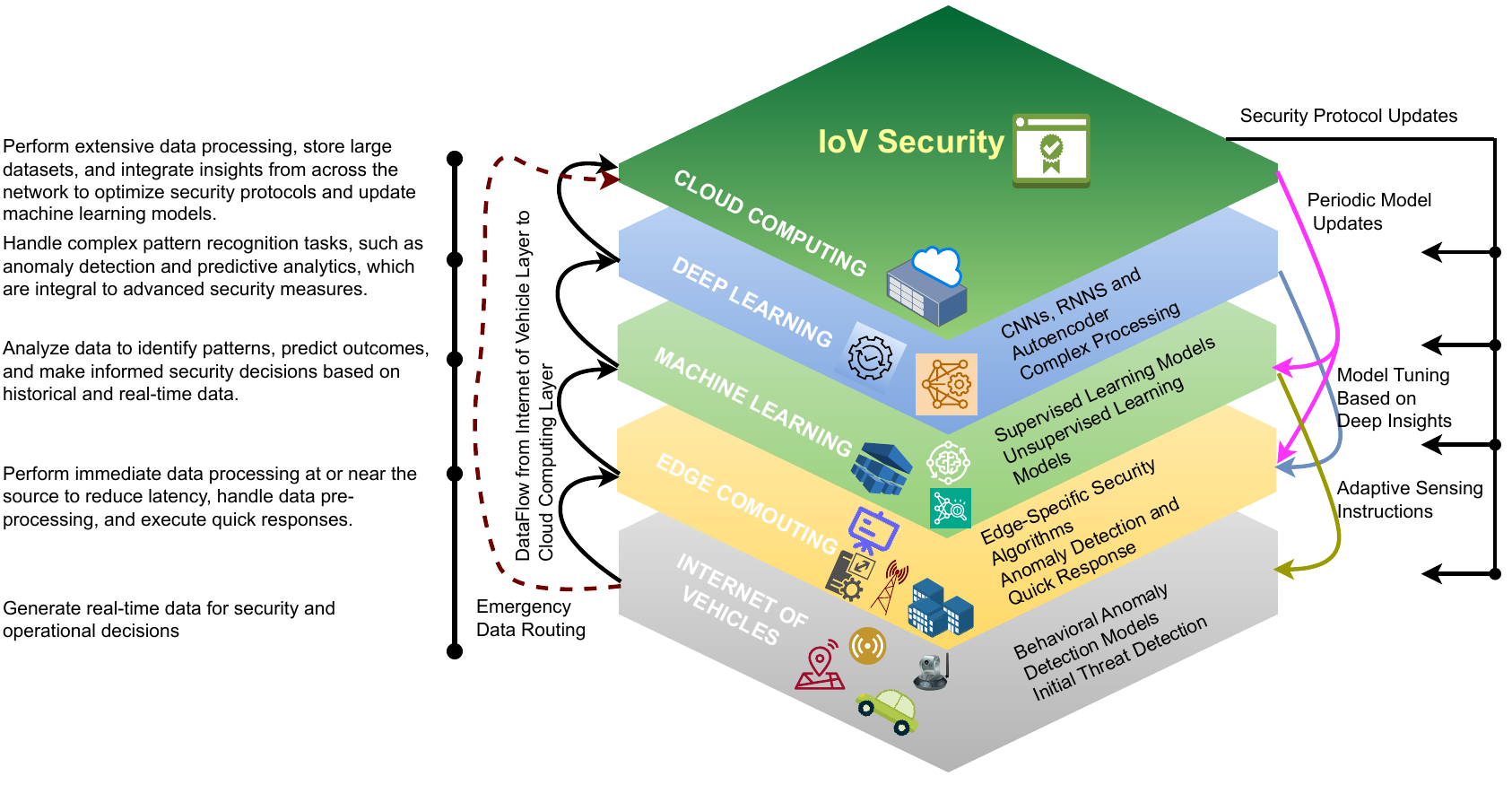}
    \caption{Integrative model of EC, ML, and DL in enhancing IoV security.}
    \label{fig:techintegration}
\end{figure*}

\section{Synergistic Integration of EC, ML, and DL for IoV Security}\label{synergistic-effects}

\subsection{Motivation for Integration}\label{sec:motivation_integration}
While individual technologies, EC, ML, and DL, have advanced IoV security, each presents distinct limitations when deployed in isolation. EC enables low-latency inference by relocating computation to OBUs and RSUs, yet its limited compute power restricts analytic depth~\cite{Xu2021Secure,Hou2020Reliable}. ML models improve detection accuracy, particularly for known threats, but struggle with zero-day attacks and adversarial inputs~\cite{Bilal2024RobustIoV,Gao2018A}. DL excels at capturing high-dimensional patterns and temporal anomalies using architectures like LSTM autoencoders and GANs~\cite{Nie2020Data-Driven,Ashraf2020Novel}, yet these gains are bounded by energy and hardware constraints in vehicular settings.

A combined architecture aligns these technologies to offset their individual constraints. EC provides the latency guarantees required for V2X safety operations, ML supports fast, adaptive policy adjustments at the edge, and DL enables deep threat insight through advanced sequence modeling and generative training. As illustrated in Figure~\ref{fig:techintegration}, this fusion creates a tiered defense stack that localizes response, enhances detection granularity, and supports real time retraining and mitigation workflows. The following subsections explore this integration framework, its benefits, and its operational challenges.
\subsection{System Architecture and Integration Stack}
\label{sec:tech_integration}

Figure~\ref{fig:techintegration} illustrates a four-tier IoV security architecture that integrates EC, ML, and DL into a unified, adaptive defense framework. Each tier contributes complementary strengths, supporting a scalable pipeline from real-time filtering to global orchestration.

At the \textit{edge tier}, OBUs and RSUs execute lightweight anomaly detection, using threshold-based filters and statistical checks to isolate malformed packets or abnormal signal patterns within sub-100\,ms latency windows~\cite{Cui2020A}. These initial responses are enhanced at the ML tier, where supervised, unsupervised, and reinforcement models refine alerts and adapt policies in response to traffic dynamics, achieving over 90\% precision on known threats~\cite{Gao2018A,Zeng2022APD:}. The \textit{DL tier} processes high-dimensional inputs, such as sensor streams and log sequences, using CNNs, RNNs, and autoencoders to detect stealthy or high-complexity threats. GANs are also employed to augment training with synthetic attack traces~\cite{Nie2020Data-Driven,Ashraf2020Novel}. Finally, the orchestration tier synchronizes updates across nodes, manages FL cycles, and implements predictive countermeasures based on inferred threat trajectories~\cite{Wang2022Deep}.

This layered architecture enables distributed IoV systems to maintain low latency, high accuracy, and robust coordination. The tiered design ensures modularity and resilience, allowing each computational layer to contribute efficiently to proactive and scalable security enforcement.

\subsection{Evaluated Benefits and Security Impact of EC-ML-DL Integration}\label{sec:integrated_gains}
The combined use of EC, ML, and DL enables IoV security systems to achieve substantial improvements in detection accuracy, responsiveness, bandwidth efficiency, and system scalability. By distributing inference across OBUs, RSUs, and cloud layers, integrated frameworks achieve end to end visibility and adaptivity. Empirical evaluations report that EC-based local filtering reduces sensor stream latency to under 120 ms~\cite{Verma2022FETCH:,Xu2021Secure}, while ML classifiers detect flow anomalies with 95--99\% precision~\cite{Gao2018A,Bilal2024RobustIoV}. DL layers such as CNNs and autoencoders further enhance detection of zero-day attacks, achieving over 99\% accuracy on benchmark datasets with diverse threat models~\cite{Xu2023Safe:,Wang2020Deep}. Beyond accuracy, integration leads to tangible performance gains. Round-trip latency is reduced from over 500 ms to approximately 115 ms by localizing inference at the edge~\cite{Bhatia2022Improved,Hou2020Reliable}, and power-efficient implementations, such as model pruning and quantization, achieve up to 60$\times$ energy savings, preserving EV battery range~\cite{Veeramanikandan2020Data}, as detailed in Table~\ref{tab:comparison_metrics_iov_full}. Scalability benefits emerge from FL and distributed model updates, which eliminate centralized bottlenecks while complying with data privacy mandates~\cite{Chen2019Deep,yang2022edge}. Additionally, blockchain-anchored logs enforce tamper resistance and prepare the system for post-quantum cryptographic resilience~\cite{Bukhari2024,Zhu2020Two}.

Proactive defense is another key advantage. DL-based time series models forecast intrusion surges and trigger preemptive mitigation, such as access throttling or node isolation, reducing response delays and improving fleet-wide resilience~\cite{Oseni2023An}. Orchestrated playbooks automate these actions, minimizing mean time to respond and increasing overall security readiness~\cite{Elsayed2020PredictDeep:}. Collectively, these outcomes demonstrate that EC-ML-DL integration is not merely additive but synergistic, achieving sub-second reaction times, privacy preservation, and high detection accuracy that standalone techniques cannot deliver in isolation.

\begin{table*}[!t]
\centering
\caption{Key metrics Before vs.\ After EC+ML/DL integration in IoV.}
\label{tab:comparison_metrics_iov_full}
\fontsize{6pt}{7.2pt}\selectfont

\begin{tabularx}{\linewidth}{%
  >{\hsize=0.3\hsize\raggedright\arraybackslash}X
  >{\hsize=0.7\hsize\raggedright\arraybackslash}X
  >{\hsize=1.5\hsize\centering\arraybackslash}X
  >{\hsize=1.5\hsize\centering\arraybackslash}X
}
\toprule
\textbf{Refs} & \textbf{Metric} & \textbf{Pre (Cloud Only)} & \textbf{Post (Edge + ML/DL)} \\
\hline
\cite{Verma2022FETCH:}  
& Response time (ms)  
& $>500$~ms  
& $\approx115$~ms \\ \hline

\cite{Verma2022FETCH:}  
& Power (W)  
& High, continuous TX  
& $\downarrow60\times$ (local processing) \\ \hline

\cite{abeshu2018deep}  
& Accuracy (\%)  
& $\approx95.2$\%  
& $\approx99.2$\% \\ \hline

\cite{Tuli2019HealthFog:}  
& Scalability  
& Poor, centralized  
& High, distributed \\ \hline

\cite{Merone2022A}  
& Adaptability  
& Manual tuning  
& Automatic reconfig. \\ \hline

\cite{Singh2022Machine-Learning-Assisted}  
& Integration complexity  
& Low; simple setup  
& High; secure sync $\checkmark$ \\ 
\bottomrule
\end{tabularx}
\end{table*}

\begin{table*}[!t]
\centering
\caption{Key barriers in IoV integration and concise solutions.}
 \label{tab:challenges_solutions_iov}
\fontsize{6pt}{7.2pt}\selectfont

\begin{tabularx}{\linewidth}{%
  >{\hsize=0.4\hsize\raggedright\arraybackslash}X
  >{\hsize=0.6\hsize\raggedright\arraybackslash}X
  >{\hsize=2.0\hsize\raggedright\arraybackslash}X
}
\toprule
\textbf{Refs} & \textbf{Barrier} & \textbf{Solutions \& Notes} \\
\hline
\cite{Yu2021EC-SAGINs:,Sacco2020An,Grover2021Edge,Cheng2019Space/Aerial-Assisted}  
& Interoperability  
& Define unified protocols; test across platforms; ensure backward compatibility \\ \hline

\cite{Popoola2021Federated,Shuvo2023Efficient,Resende2021TIP4.0:,Liu2020Resource}  
& Maintenance \& Updates  
& Automate health checks and patching; schedule regular software/hardware audits \\ \hline

\cite{Sullivan2019EU,Markopoulou2019The,Mantelero2020The}  
& Regulatory  
& Align with GDPR, CCPA, NIS, NIST; embed compliance checks in workflows \\ \hline

\cite{Dai2019Industrial,Garg2021Security,Ferrag2022Edge-IIoTset:,He2021Blockchain-Based}  
& Industry Standards  
& Standardize data formats \& security specs; adopt ISO/SAE 21434; liaise with regulatory bodies \\ 
\bottomrule
\end{tabularx}
\end{table*}

\subsection{Integration Challenges and Design Considerations}
\label{sec:int_integration_challenges}

While the integration of EC, ML, and DL enhances IoV security, it also introduces complex operational and architectural challenges. These challenges span system interoperability, lifecycle maintenance, and regulatory compliance, all of which must be addressed to ensure secure, scalable, and standards-aligned deployments. Table~\ref{tab:challenges_solutions_iov} summarizes key barriers and mitigation strategies reported in recent literature.

\subsubsection{System Interoperability}
Integrated IoV systems often suffer from mismatches in data granularity and representation. Edge-layer devices emit low-dimensional metrics, while DL models require structured, high-dimensional tensors. Without standardized transformation pipelines, these format mismatches can lead to false positives or dropped alerts~\cite{Sacco2020An,Grover2021Edge}. Furthermore, legacy communication protocols such as CAN and DSRC complicate AI-driven integration, particularly when aligning inputs across heterogeneous modules~\cite{Cheng2019Space/Aerial-Assisted}. To bridge this gap, recent studies propose adaptive feature conversion pipelines and standard-compliant APIs for secure model interfacing.

\subsubsection{Model Maintenance and Lifecycle Management}
Security models must continuously adapt to shifting traffic patterns and evolving attack vectors. However, maintaining model relevance across distributed nodes introduces risks related to update validation, backward compatibility, and runtime integrity. Liu et al.~\cite{Liu2020Resource} and Resende et al.~\cite{Resende2021TIP4.0:} emphasize that poorly synchronized updates or aging hardware can degrade detection accuracy or interrupt safety-critical functions. Emerging DevSecOps pipelines push signed updates over-the-air, while local health monitors detect degraded or misaligned nodes~\cite{Ferrag2021Federated,Shuvo2023Efficient}. FL complements these approaches by enabling local retraining, reducing the reliance on centralized data transfer~\cite{Popoola2021Federated}.

\subsubsection{Regulatory Compliance and Data Governance}
Integrated AI-driven security must operate within the constraints of global data protection regulations such as GDPR, CCPA, and UNECE WP.29~\cite{Sullivan2019EU,Markopoulou2019The,Mantelero2020The,GDPR_2016,CCPA_2018}. These mandates require traceability, auditability, and minimal exposure of personal data, particularly for edge-to-cloud data pipelines. Solutions such as differential privacy, encrypted model sharing, and federated inference workflows have been proposed to satisfy legal constraints without undermining analytical accuracy~\cite{Yu2021EC-SAGINs:}. Aligning these strategies with evolving regulatory frameworks remains an ongoing research and engineering challenge.

Together, these design considerations underscore the need for co-optimized software-hardware stacks, lifecycle-aware update systems, and privacy-centric architectures that can operate reliably under real-world IoV constraints.

\subsection{Standardization and Regulatory Compliance in IoV Security}\label{sec:standards_compliance}
Integrating edge computing, machine learning, and deep learning into IoV security frameworks requires adherence to evolving regulatory mandates and technical standards. Without standardization, deployments risk interoperability failures, fragmented data practices, and non-compliance with global cybersecurity legislation. This subsection synthesizes industry-aligned efforts and regulatory mechanisms that shape integrated IoV security.

\subsubsection{Interoperability and Framework Alignment}
IoV ecosystems incorporate diverse components, e.g., OBUs, RSUs, VANETs, and AI-based models, each operating under varied communication and security protocols. Standards such as ISO/SAE 21434 provide guidance on lifecycle cybersecurity engineering, promoting cohesion across devices, interfaces, and vendors~\cite{Garg2021Security,Choi2017Special}. Protocol-level standards like IEEE 802.11p ensure secure and low-latency V2V/V2I communication, which is essential for harmonizing edge inference with AI-driven control loops~\cite{jiang2008ieee}.

\subsubsection{Privacy and Legal Compliance}
IoV platforms must align with data protection laws including the GDPR, CCPA, and UNECE WP.29, which mandate strict privacy controls, lawful data usage, and security by design~\cite{Sullivan2019EU,Markopoulou2019The,Mantelero2020The,GDPR_2016}. ML and DL frameworks that process personally identifiable vehicle data must ensure data minimization and traceability. FL, blockchain anchoring, and encrypted inference have emerged as viable compliance tools, preserving functionality while meeting legal requirements~\cite{Ferrag2022Edge-IIoTset:,Yu2021EC-SAGINs:}.

\subsubsection{Security and Real-Time Processing Standards}
Cybersecurity risk management frameworks such as ISO 27001 and NIST SP 800-53 outline best practices for ensuring confidentiality, integrity, and availability in AI-assisted systems~\cite{ISO_IEC_27001_2022,NIST_SP_800-53r5_2020}. These standards support the development of certified DL-based intrusion detection and anomaly response tools. In parallel, emerging standards like IEEE P2418.5 and ISO/TC 307 define blockchain integration protocols, facilitating decentralized trust and tamper-evident logging in IoV contexts~\cite{IEEEP2418,ISO_TC_307_2024}.

\subsubsection{Toward AI-Specific IoV Standards}
As AI becomes integral to vehicular security, future standards must address explainability, adversarial robustness, and trust in autonomous models. The literature emphasizes the need for domain-specific compliance frameworks that bridge AI lifecycle governance with real-time vehicular constraints~\cite{Mantelero2020The,Dai2019Industrial}. Table~\ref{tab:comparison_metrics_iov_full} and Table~\ref{tab:challenges_solutions_iov} highlight the interplay between performance, privacy, and standard conformance across current deployments.

\begin{table*}[!t]
\centering
\caption{Edge-computing case studies in IoV security: Entities, tech, and core impacts.}
\label{tab:company_tech_iov_EC}
\fontsize{6pt}{7.2pt}\selectfont

\begin{tabularx}{\linewidth}{%
  >{\hsize=0.1\hsize\raggedright\arraybackslash}X
  >{\hsize=0.7\hsize\raggedright\arraybackslash}X
  >{\hsize=0.9\hsize\raggedright\arraybackslash}X
  >{\hsize=2.3\hsize\raggedright\arraybackslash}X
}
\toprule
\textbf{Ref} & \textbf{Entity} & \textbf{Tech} & \textbf{Impact \& Applications} \\
\hline
\cite{porsche2019}  
& Porsche  
& FogHorn EC  
& $\checkmark$ Multi-factor auth; traffic optimization; predictive maintenance; V2H infotainment \\ \hline

\cite{volvo2021}  
& Volvo  
& EC-driven collision avoidance  
& $\checkmark$ Real-time safety decisions; ``Safe Space'' accident prevention \\ \hline

\cite{lenovo2021}  
& Barcelona (Smart City)  
& IoT+EC  
& $\checkmark$ Congestion reduction; adaptive traffic lights \\ \hline

\cite{citiesforum2023}  
& Shanghai (Smart City)  
& Intersection EC  
& $\checkmark$ Smooth traffic flow; real-time signal control \\ 
\bottomrule
\end{tabularx}
\end{table*}

\section{Deployment Experiences and Lessons from EC, ML, and DL in IoV}
\label{case-studies}

\subsection{Use Cases of EC, ML, and DL Technologies}\label{sec:deployment_overview}
Edge computing, machine learning, and deep learning have transitioned from theoretical constructs to operational pillars in modern IoV deployments. Table~\ref{tab:company_tech_iov_EC} and Table~\ref{tab:DL_application_domains_iov} summarize representative implementations across automotive and smart city contexts.

In \textit{Vehicle-Centric Deployments}, Porsche and FogHorn embedded EC modules for offline multimodal biometric authentication, enabling secure, keyless entry and predictive diagnostics without relying on cloud connectivity~\cite{porsche2019}. Volvo’s Safe Space system integrated onboard EC to process real-time sensor fusion from LiDAR, radar, and cameras, achieving automated braking in under 50 ms~\cite{volvo2021}. Similarly, autonomous vehicle fleets trialed CNN-LSTM models for real-time obstacle detection and spoofed signal mitigation, with companies like Waymo reporting notable safety enhancements~\cite{waymo2020,frost2023}. \textit{ML-based intrusion detection and predictive maintenance systems} have also entered production, with Yang et al.~\cite{Yang2020Machine} and Adi et al.~\cite{Adi2020Machine} demonstrating adaptive retraining to address zero-day threats and forecast mechanical failures. Additional use cases include traffic flow optimization~\cite{Sun2018Application} and automated vehicle isolation in response to detected intrusions~\cite{Liu2018A}. A broader summary of these deployments and their outcomes is presented in Table~\ref{tab:ml_case_studies_iov_security}.

\begin{table*}[!t]
\centering
\caption{DL application domains in IoV security.}
\label{tab:DL_application_domains_iov}
\fontsize{6pt}{7.2pt}\selectfont
\begin{tabularx}{\linewidth}{%
  >{\hsize=0.5\hsize\raggedright\arraybackslash}X
  >{\hsize=0.7\hsize\raggedright\arraybackslash}X
  >{\hsize=1.3\hsize\raggedright\arraybackslash}X
  >{\hsize=1.5\hsize\raggedright\arraybackslash}X
}
\toprule
\textbf{Ref} & \textbf{Domain} & \textbf{Implementation} & \textbf{Benefit} \\
\hline
\cite{Grover2021Edge,Nie2020Data-Driven,Oseni2023An}
& Network Security
& CNN/LSTM monitoring
& Ensures comms integrity \\ \hline

\cite{frost2023,waymo2020}
& Autonomous Vehicles
& Sensor‐data DL models
& Reduces accidents and breaches \\ \hline

\cite{ptvgroup2024}
& Traffic Mgmt
& Flow‐analysis networks
& Minimizes disruptions \\ \hline

\cite{Fathy2022Integrating,Sikora2020Artificial,Wang2018Vehicle}
& Surveillance
& Video‐analytics CNNs
& Improves incident response \\ 
\bottomrule
\end{tabularx}
\end{table*}

\begin{table*}[!t]
\centering
\caption{ML case studies in IoV security: approach, outcome, and lesson.}
\label{tab:ml_case_studies_iov_security}
\fontsize{6pt}{7.2pt}\selectfont
\begin{tabularx}{\linewidth}{%
  >{\hsize=0.2\hsize\raggedright\arraybackslash}X
  >{\hsize=1.1\hsize\raggedright\arraybackslash}X
  >{\hsize=1.2\hsize\raggedright\arraybackslash}X
  >{\hsize=1.5\hsize\raggedright\arraybackslash}X
}
\toprule
\textbf{Ref} & \textbf{Approach} & \textbf{Outcome} & \textbf{Lesson} \\
\hline
\cite{Yang2020Machine}
& Continuous learning framework
& Real‐time mitigation of simulated attacks
& Ongoing model adaptation is critical \\ \hline

\cite{Adi2020Machine}
& Sensor‐data predictive modeling
& Enhanced safety; reduced downtime
& ML adds operational efficiency beyond security \\ \hline

\cite{Sun2018Application}
& Real‐time traffic‐pattern analysis
& Optimized flow; lower delays
& ML supports broader IoV ecosystem management \\ \hline

\cite{Liu2018A}
& Automated detection and isolation
& Contained threat spread swiftly
& Automation dramatically improves response time \\ 
\bottomrule
\end{tabularx}
\end{table*}

\textit{Urban and infrastructure-focused deployments} extend these innovations to smart cities. Barcelona, Singapore, and Wuxi have implemented intersection-based EC nodes equipped with lightweight ML models for dynamic traffic signal control, reducing congestion and improving autonomous coordination~\cite{lenovo2021,citiesforum2023,app11209680}. In parallel, cities like Vienna and Rome use DL-driven surveillance to detect suspicious gatherings and road anomalies in real time~\cite{ptvgroup2024}. For network-layer protection, CNN/LSTM-based IDS platforms have been applied to monitor CAN traffic and V2X logs, achieving high accuracy in detecting anomalous behavior~\cite{Grover2021Edge,Nie2020Data-Driven}. \emph{Integrated EC-ML-DL systems} further demonstrate compound benefits. In one example, EC was combined with lightweight ML to enable in-vehicle intrusion response, cutting cyberattack success rates in fleet trials~\cite{Zhang2019Mobile}. Meanwhile, EC-enhanced RNNs and RL models were deployed in urban intersections to secure V2X channels and defend against coordinated cyberattacks, delivering sub-100\,ms response times and improved threat isolation~\cite{Wang2020Deep,Liu2019Edge}.

These deployments illustrate the growing maturity of intelligent, distributed IoV security infrastructure. They also provide empirical validation for the capabilities and constraints of EC, ML, and DL when applied to heterogeneous vehicular and urban environments.

\subsection{Lessons Learned from Real-World Deployments}\label{sec:deployment_lessons}
Deployment experiences across edge-enabled, ML-driven, and DL-enhanced IoV systems reveal several recurring design principles and operational challenges that inform future architecture development and standardization.

First, low-latency edge processing emerges as a foundational enabler for real-time threat detection and system responsiveness. Across vehicle and infrastructure contexts, deployments consistently achieve sub-100\,ms reaction times through on-device ML filtering and localized DL inference~\cite{volvo2021,Grover2021Edge,Wang2020Deep}. This shift toward edge-centric analytics also alleviates cloud dependency, reduces bandwidth usage, and minimizes exposure of sensitive data, supporting both security and privacy objectives.

Second, continuous adaptation is critical to sustaining protection in dynamic environments. Many successful deployments incorporate model retraining, automated update pipelines, or FL frameworks to handle evolving traffic patterns and adversarial behavior~\cite{Yang2020Machine,Garg2021Security}. Techniques such as pruning and quantization allow ML and DL models to operate efficiently on constrained edge hardware without compromising inference performance~\cite{Shuvo2023Efficient}.

Third, deployment at scale surfaces interoperability and governance challenges. Heterogeneous ECUs, RSUs, and cloud systems require standardized data formats, communication protocols, and secure APIs to ensure seamless integration and coordinated defense~\cite{Zhang2019Mobile,Alonso2020Deep}. Privacy-preserving analytics and end-to-end encryption are also essential for maintaining compliance with data protection regulations and retaining user trust~\cite{Tian2020A}.

These insights underscore the need for co-designed IoV security solutions that unify distributed intelligence with practical deployment constraints. Real-world evidence affirms the potential of EC, ML, and DL integration to deliver resilient, efficient, and adaptive protection, but also highlights the importance of modularity, explainability, and regulatory alignment.

\begin{table*}[!t]
\centering
\caption{Research areas in AI‐powered IoV security: condensed challenges and key questions.}
\label{tab:research_gaps_future_directions}
\fontsize{6pt}{7.2pt}\selectfont

\begin{tabularx}{\linewidth}{%
  >{\hsize=0.35\hsize\raggedright\arraybackslash}X
  >{\hsize=0.9\hsize\raggedright\arraybackslash}X
  >{\hsize=1.75\hsize\raggedright\arraybackslash}X
}
\toprule
\textbf{Area} & \textbf{Core Challenges} & \textbf{Key Research Questions} \\
\hline
Deployment & Compute efficiency; adaptive inference; edge-cloud orchestration & 
$\triangleright$ Can one compression policy support diverse sensor streams with sub-100\,ms latency?\newline
$\triangleright$ How to partition models dynamically across OBUs, RSUs, and cloud?\newline
$\triangleright$ Is event-driven inference viable for multi-modal inputs? \\ \hline

Explainability & Regulatory compliance; runtime explainability; secure auditing & 
$\triangleright$ How to deliver XAI insights under 5\,ms latency and minimal compute cost?\newline
$\triangleright$ How can root-cause analysis improve forensic reliability?\newline
$\triangleright$ How to certify XAI-enhanced IDSs without exposing proprietary models?\\ \hline

Adversarial Risks & Perturbation robustness; privacy leaks; attack resilience & 
$\triangleright$ Which certified defenses scale to real-time IoV settings?\newline
$\triangleright$ How to preserve privacy in non-IID federated training?\newline
$\triangleright$ What adversarial training schedules fit 2\,GB edge constraints?\\ \hline

Crypto Security & Quantum-safe handshakes; encrypted inference; chain bloat & 
$\triangleright$ Can hybrid post-quantum crypto meet $<$5\,ms V2X latency targets?\newline
$\triangleright$ How practical is encrypted CNN inference on Cortex-class MCUs?\newline
$\triangleright$ How to balance backward compatibility with forward secrecy?\\ \hline

Threat Sharing & Cross-OEM coordination; data siloing; trusted automation & 
$\triangleright$ Can smart contract-based intel sharing meet $<$50\,ms latency at scale?\newline
$\triangleright$ How to incentivize early threat reporting in federated settings?\newline
$\triangleright$ How to aggregate threat insights without raw data exposure?\\ \hline

Future Architecture & Cross-layer orchestration; energy limits; quantum resilience & 
$\triangleright$ How to secure DL/ML/crypto tasks across fog/edge/cloud in real time?\newline
$\triangleright$ Can RL optimize latency-energy tradeoffs dynamically?\newline
$\triangleright$ How to meet quantum safety within a 10\,W edge power envelope?\\
\bottomrule
\end{tabularx}
\end{table*}

\section{Future Directions and Research Opportunities}\label{PredictionsandResearchOpportunities}
AI-enabled IoV systems unlock new capabilities but broaden the attack surface. Advancing security now hinges on three fronts: (i) sub-second threat detection on constrained edge nodes, (ii) privacy-preserving analytics compliant with GDPR/WP.29, and (iii) scalable, explainable, and quantum-resilient architectures. Table~\ref{tab:research_gaps_future_directions} outlines open challenges, from balancing accuracy and latency to hardening federated models against poisoning, that will define next-gen IoV security.

\subsection{Resource-Aware AI at the Edge}\label{sec:future_resource}

\indent\bolditalic{Current landscape.}~Edge NPUs can deliver 4-20 TOPS at under 5 W, but full-precision models like ResNet still overwhelm mid-tier OBUs in dense traffic. Compression pipelines (e.g., AMC, AutoSlim) prune up to 90\% of parameters, but often yield just 30\% latency gains.

\bolditalic{Emerging directions.}~Innovations span model partitioning, event-driven compute, and self-supervision. \emph{Federated splitting} dynamically assigns early layers to OBUs, heavier extractors to RSUs, and aggregation to the cloud, adapting to RF conditions. \emph{Event-driven inference} activates only upon salient input, cutting energy use 5-10$\times$. \emph{Self-supervised distillation} uses contrastive or masked objectives to compress models without labels, ideal for privacy-conscious IoV contexts.

\bolditalic{Open question.}~Can one policy achieve 98\% accuracy, $<$100\,ms latency, and 1 TOPS/W across 30 fps video and sporadic LiDAR?

\subsection{Explainable \& Auditable Security AI}\label{sec:future_xai}
\bolditalic{Problem.}~A blocked brake-assist due to an opaque IDS verdict could be fatal. Regulations now demand ``meaningful explanations” (UNECE WP.29 §7, ISO/SAE 21434 §9).

\bolditalic{Emerging directions.}~\emph{Edge-native SHAP/LIME} seeks $<$5\,ms latency and $<$3\,\% CPU overhead. \emph{Counterfactual root-cause analysis} autogenerates ``nearest safe sample'' explanations for anomalies, easing forensics. \emph{Blockchain-anchored model provenance} hashes updates and triggers automatic rollbacks via smart contracts when KPIs drift or bias increases.

\bolditalic{Open question.}~How can we certify XAI-enabled IDSs without disclosing proprietary weights while satisfying auditors?

\subsection{Robustness to Adversarial \& Privacy Attacks}\label{sec:future_adv}
\bolditalic{Threats.}~Real-world patch attacks fool sign classifiers at highway speeds; CAN-bus poisoning skews anomaly scores; membership inference leaks driver habits.

\bolditalic{Emerging directions.}~\emph{Certified defenses} (e.g., randomized smoothing, Lipschitz bounds) offer robustness up to perturbation $\varepsilon$. \emph{Differentially private FL} adds calibrated noise for privacy under non-IID conditions. \emph{Edge-practical adversarial training} integrates curriculum attacks during training, boosting resilience with $<$20\,\% overhead.

\bolditalic{Open question.}~Which lightweight training schedule retains $\geq 95\%$ AUC while fitting within 2~GB OBU flash?

\subsection{Next-Generation Cryptography}\label{sec:future_crypto}
\bolditalic{Limitations.}~RSA-2048 adds $\sim$33\,ms handshake; ECC-256 meets latency needs but lacks post-quantum security.

\bolditalic{Emerging directions.}~\emph{Lattice-based Ring-LWE signatures} compress to $\sim$1\,KB, verified in 2-3\,ms on Cortex-A55. \emph{Hybrid cryptography} combines ECQV credentials with post-quantum key encapsulation. \emph{Efficient homomorphic inference} (CKKS, TFHE) supports encrypted CNNs at $\leq 2\times$ latency overhead for low-res sensors.

\bolditalic{Open question.}~Can hybrid post-quantum + symmetric schemes enable V2X handshakes $<$5\,ms on 32-bit MCUs?

\subsection{Federated Threat Intelligence Ecosystems}\label{sec:future_threatshare}
\bolditalic{Problem.}~Central SOCs strain under 50\,TB/day fleet telemetry; siloed feeds hinder cross-OEM awareness.

\bolditalic{Emerging directions.}~\emph{STIX/TAXII over blockchain} offers standardized, immutable threat intel with on-chain incentives. \emph{Smart contract patch pipelines} automate revocations and hotfixes once consensus is met. \emph{Hierarchical federated IDSs} preserve OEM data while sharing distilled threat vectors.

\bolditalic{Open question.}~How can threat intel smart contracts maintain $<$50\,ms latency across 1M vehicles without bloating the chain?

\subsection{Cross-Layer \& Quantum-Resilient Architectures}\label{sec:future_arch}
\bolditalic{Vision.}~A dynamic orchestration plane that:~\emph{(i)}~assigns DL, ML, and crypto tasks across edge/fog/cloud in real time; \emph{(ii)}~uses RL to balance latency, energy, and link quality; \emph{(iii)}~secures end-to-end with post-quantum cryptography.

\bolditalic{Grand challenge.}~Achieve a sub-10\,ms defense loop that is robust to $\varepsilon$-bounded attacks, GDPR-compliant, and quantum-safe, all within a 10~W edge envelope.

Meeting these challenges will shape a future of resilient, explainable, and quantum-secure IoV security.

\section{Conclusion}\label{sec:Conclusion}
In this survey, we have examined how EC, ML, and DL technologies, individually and in concert, can address the unique security challenges of the IoV. EC brings critical low latency, localized processing that shrinks the attack surface and enables real time anomaly filtering. ML contributes adaptive, data driven detection of both known and novel threats, while DL unlocks high-dimensional pattern recognition for subtle or zero-day attack signatures.  Our comparative analysis demonstrated that no single technology suffices: only a layered EC-ML-DL stack can deliver sub-10 ms response, > 99\% detection accuracy, and scalable, privacy preserving operation. Real world case studies from automotive OEMs and smart cities confirm the feasibility and benefits of these integrations, keyless edge authentication at Porsche, millisecond scale collision avoidance at Volvo, adaptive traffic control in Barcelona, and fleet wide intrusion shields in autonomous vehicles.  Lessons learned underscore the importance of robust data governance, interoperable protocols, and continuous model adaptation to keep pace with evolving threats. Additionally, IoV security must overcome resource constraints, ensure explainability, and achieve resilience against adversarial and quantum era attacks.  Promising research avenues include federated and event driven edge learning, post quantum cryptography tailored to V2X, and blockchain anchored audit trails for model integrity.  By charting these directions, we aim to provide researchers and practitioners with a roadmap for building the next generation of secure, efficient, and intelligent IoV systems, capable of safeguarding connected transportation ecosystems.  

\balance
\bibliographystyle{ACM-Reference-Format}
\bibliography{References_CSUR_v2}
\end{document}